# A complete phase-field fracture model for brittle materials subjected to thermal shocks

Bo Zeng and John E. Dolbow

Department of Mechanical Engineering and Materials Science, Duke University, Durham, NC 27708, USA

Brittle materials subjected to thermal shocks experience strong temperature gradients that in turn give rise to mechanical stresses that can be large enough to induce fracture. This work presents a complete model for phase-field fracture for coupled thermo-mechanical problems, wherein the bulk material properties, the material strength, and the fracture toughness are specified independently. The capabilities of the model are assessed across a wide span of scenarios in thermo-mechanical fracture, from the propagation of large pre-existing cracks to crack nucleation under spatially uniform states of stress. In particular, we revisit the controlled quenching of glass plates, and demonstrate how the model captures experimentally observed crack patterns across a range of thermal loads. Ceramic disks subjected to infrared radiation are also examined, with the model reproducing both straight cracks in notched specimens and branching in intact specimens. Finally, ceramic pellets subjected to rapid power pulses are examined, with the model explaining experimental transitions from intact to fractured pellets and the important role of material strength. The results demonstrate that the complete phase-field model unifies the treatment of distinct fracture scenarios under thermal shock, surpassing classical approaches and enabling more reliable prediction of brittle fracture in extreme environments.

**Keywords:** phase-field for fracture, material strength, nucleation, thermally induced fracture, dynamic branching

## 1 Introduction

Understanding and predicting the behavior of materials subjected to large temperature gradients has been a fundamental pursuit of scientists and engineers for over a century. From ancient pottery shattered by sudden heating to modern turbine blades, furnace linings, and thermal barrier coatings, material fracture under thermal shock remains a critical concern across many industries. During a thermal shock, for example, rapid temperature changes induce large mechanical stresses, causing materials to crack unexpectedly. Despite remarkable technological advances (Manson 1954; Hasselman 1969; Meng et al. 2024), accurately predicting how brittle materials respond to extreme heating and sudden cooling remains an open scientific challenge. To address this challenge, several key factors governing categories of fracture processes have been identified, contributing to the fundamentals of predictive and reliable models.

The long-standing theoretical framework for thermal shock fracture in brittle materials, developed by Manson (1954) and Hasselman (1969) among others, is rooted in linear elastic fracture mechanics (LEFM) and characterizes fracture susceptibility through thermal shock resistance parameters. These classical results have informed materials design for decades, but they assume the existence of pre-existing cracks and are not designed to describe crack nucleation or complex evolving crack paths. For example, methods based on LEFM for thermal shock problems tend to rely on ad-hoc criteria to model phenomena such as crack branching (Chiaramonte et al. 2020).

Phase-field regularizations (Bourdin et al. 2000) of the variational theory of brittle fracture of Francfort and Marigo (1998) provide a more general alternative. In these models, the sharp crack surface is replaced by a diffuse field governed by a regularized variational principle. The approach naturally accommodates complex crack paths, branching, and coalescence without explicit tracking, and has been extended to thermo-mechanical settings by a number of groups (Corson et al. 2009; Kilic and Madenci 2009; Menouillard and Belytschko 2011; Chu et al. 2017; Xu et al. 2018; Wang et al. 2018; Svolos et al. 2020; Mandal et al. 2021; Li et al. 2023; Pan et al. 2024; Chen et al. 2024). These thermally coupled formulations have demonstrated the ability to capture a variety of crack patterns in benchmark problems involving quenched glass plates and heated ceramic specimens, and have examined topics such as the role of thermal conductivity degradation across crack surfaces (Svolos et al. 2020) and the use of advanced discretization strategies (Li et al. 2023; Chen et al. 2024).

Despite this progress, all of the aforementioned thermally coupled phase-field formulations inherit a fundamental physical limitation of classical variational phase-field models: they do not treat material strength as an independent constitutive property. In classical variational phase-field models[1], the uniaxial tensile strength scales as $\sigma_{\text{ts}} \sim \sqrt{G_c E/\ell}$, where $E$ is Young's modulus, $G_c$ the fracture toughness, and $\ell$ the regularization length. As such, the threshold at which fracture initiates is not an independently prescribed material property but is instead subjugate to the fracture toughness and elasticity (Lopez-Pamies et al. 2025). This conflation of nucleation and propagation criteria has a direct physical consequence: these models cannot correctly predict crack nucleation in the bulk under spatially uniform or nearly uniform stress states, nor can they capture the mediation between strength and energetics that governs transitions between distinct fracture regimes.

Building on this foundation, the complete model introduced by Kumar, Francfort, et al. (2018); Kumar et al. (2020); Kumar et al. (2022) treats the elasticity, the fracture toughness, and the material strength surface $\mathcal{F}(\boldsymbol{\sigma}) = 0$ as three fully independent macroscopic material properties. The model is termed *complete* precisely because it independently incorporates all three of these constitutive ingredients. The theory dictates that: (i) cracks nucleate and propagate only in regions where the strength surface of the material has been exceeded, a necessary but not sufficient condition; and (ii) they do so in a manner that minimizes the sum of the deformation and fracture surface energies, which provides the sufficiency condition. Importantly, the complete model does not arise from the minimization of a total energy functional. Rather, it augments the phase-field evolution equation with an external microforce whose explicit form is determined by the material's strength surface. This structural distinction from classical variational phase-field models is precisely what enables the model to correctly predict crack nucleation at the prescribed strength threshold while simultaneously recovering Griffith-type propagation for large pre-existing cracks. Importantly, the variational basis for the structure of the complete model was established by Larsen et al. (2024), placing it on solid mathematical footing.

The complete model has been validated against a wide range of fracture experiments—spanning ceramics, elastomers, and other brittle materials, across specimen geometries from smooth bars to notched plates—that cannot be reproduced by classical variational phase-field models (Kamarei et al. 2024; Kumar et al. 2024). It exhibits the valuable attribute that, unlike many phase-field models of fracture, it is insensitive to the choice of regularization length, provided it is taken to be sufficiently small. Notably, Lopez-Pamies and collaborators recently introduced a systematic series of nine benchmark problems, spanning the full range of elastic brittle fracture scenarios, specifically designed to assess the predictive capabilities of fracture models (Kamarei et al. 2026). To date, the complete model is the only fracture formulation capable of passing all nine of these benchmarks.

Of particular relevance to the present work, the complete model has already been applied to a canonical thermal shock benchmark: the rapid quenching of alumina plates studied experimentally by Jiang et al. (2012), in which a periodic array of cracks nucleates along the quenched edge and subsequently undergoes selective arrest as the cracks propagate into the interior. This problem was examined by Bourdin et al. (2014) using a classical variational phase-field model, and has since become a standard reference problem in the phase-field fracture community. In (Zeng et al. 2025), we showed that the complete model, when combined with a spatially varying strength field, reproduces the experimentally measured crack spacing and the bimodal distribution of crack lengths across the full range of quench temperatures, as well as the hierarchical arrest of cracks reported in subsequent experiments (Shao et al. 2017; Shao et al. 2018). That study also demonstrated that the same agreement is not obtained when the strength is held spatially uniform, nor with a classical variational model in which an equivalent strength perturbation is introduced through the regularization length. The present work builds directly on this foundation. Whereas Zeng et al. (2025) focused on nucleation under nearly uniform stress states, the problems examined here span a broader range of thermal shock scenarios, from the propagation of a large pre-existing crack, to nucleation from a notch, to dynamic branching and fully three-dimensional transient loading, with the aim of demonstrating that a single formulation with independently prescribed strength accounts for all of them.

[1] By classical variational phase-field models, we mean specifically those phase-field models of fracture that Γ-converge to the variational theory of brittle fracture of Francfort and Marigo (1998).

In this work, we extend the complete model of fracture to coupled thermo-mechanical problems and demonstrate that the resulting formulation has the correct physical structure to explain a set of canonical thermal shock fracture experiments. The objective is not to introduce a new computational method or to claim efficiency advantages over existing approaches. Rather, it is to show that independently specifying the material strength—alongside the elasticity and fracture toughness—is the key physical ingredient needed to correctly explain experimental observations that have resisted satisfactory explanation by prior thermally coupled phase-field formulations. We begin by considering the simplest version of thermo-mechanical coupling that encapsulates the requisite physics. In particular, thermo-mechanical loading enters the framework through the elastic strain energy via a standard thermoelastic constitutive relation. The formulation adopts a one-way thermal coupling: the temperature field drives thermal expansion and therefore the mechanical and fracture fields, but mechanical deformation does not feed back into the heat equation and fracture energy dissipation is not coupled to the thermal field. While such couplings may be relevant in other applications, they are not required to reproduce the observations examined here. The capabilities of the resulting model are assessed across three representative benchmark problems that have been examined both experimentally and computationally by previous researchers.

The paper is organized as follows. We begin, in Section 2, by introducing the complete model of phase-field fracture, along with the spatial and temporal discretization scheme. In Section 3, we demonstrate the modeling capabilities of the complete model for the three benchmark problems. Lastly, the key findings of this study are summarized in Section 4.

## 1.1 Alternative approaches

Although this manuscript focuses on a complete phase-field model for thermal shock problems, at this point it is appropriate and illuminating to comment on alternative methods. The challenge of independently prescribing the strength and fracture toughness of a material in models for fracture has been approached from several directions. The coupled criterion of Leguillon (2002) requires simultaneous satisfaction of a stress condition and an energy condition for crack initiation, and has been applied to a range of notch and interface problems (Leguillon 2002). While influential, this approach does not yet incorporate a sufficiently general definition of material strength, provides no prediction of crack path after initiation, and remains computationally demanding in three dimensions for problems with unknown crack trajectories. Indeed, the analysis of thermo-mechanical problems using the coupled criterion has been limited to cracks that are geometrically simple, see for example (Leguillon et al. 2015; Faria Ricardo et al. 2020).

More recently and within the phase-field setting, alternatives to the complete model are beginning to emerge. This is no doubt due to the recognition of the important role played by material strength. In contrast to classical variational phase-field models of fracture, these recent alternatives have been formulated to Γ-converge to *cohesive* models of fracture. This change in construction is what has permitted them with a means to incorporate more general strength surfaces, provided they are convex. Examples of these models include the work of Vicentini et al. (2024) and Bourdin et al. (2025). To date, these models have been used to simulate relatively simple, purely mechanical fracture problems. The extent to which they might satisfy the nine circles of elastic brittle fracture or be employed for a more general class of fracture problems remains to be seen.

Finally, we mention the use of peridynamics to study thermal shock problems. Peridynamics is a nonlocal reformulation of continuum mechanics, introduced by Silling (2000), in which the equations of motion are cast as integro-differential equations that allow material points to interact over a finite horizon. This makes computational methods based on the theory simpler to use for fracture problems, as cracks nucleate and propagate without requiring any special treatment of discontinuities. Of course, simplicity of use is one thing, predictive power quite another. A key question with peridynamics concerns the role of the horizon length, and how it should be selected. Along these lines and with reference to thermal shock problems, the recent work of Xu et al. (2018) is perhaps illustrative. Indeed, their results for progressive quenching simulations show a clear dependence of the resulting fracture patterns on the particular choice of horizon length, see (Xu et al. 2018, Figure 13). This suggests that the horizon length is not so much a material property but rather something that needs to be calibrated to individual experiments, significantly diminishing the model's predictive capabilities.

# 2 Problem Formulation

In this section, we introduce the physical models employed in this study. We begin by sequentially describing the thermal, mechanical, and fracture models, followed by a discussion of the discretization strategy and the accompanying implementational details for solving the coupled system of nonlinear equations.

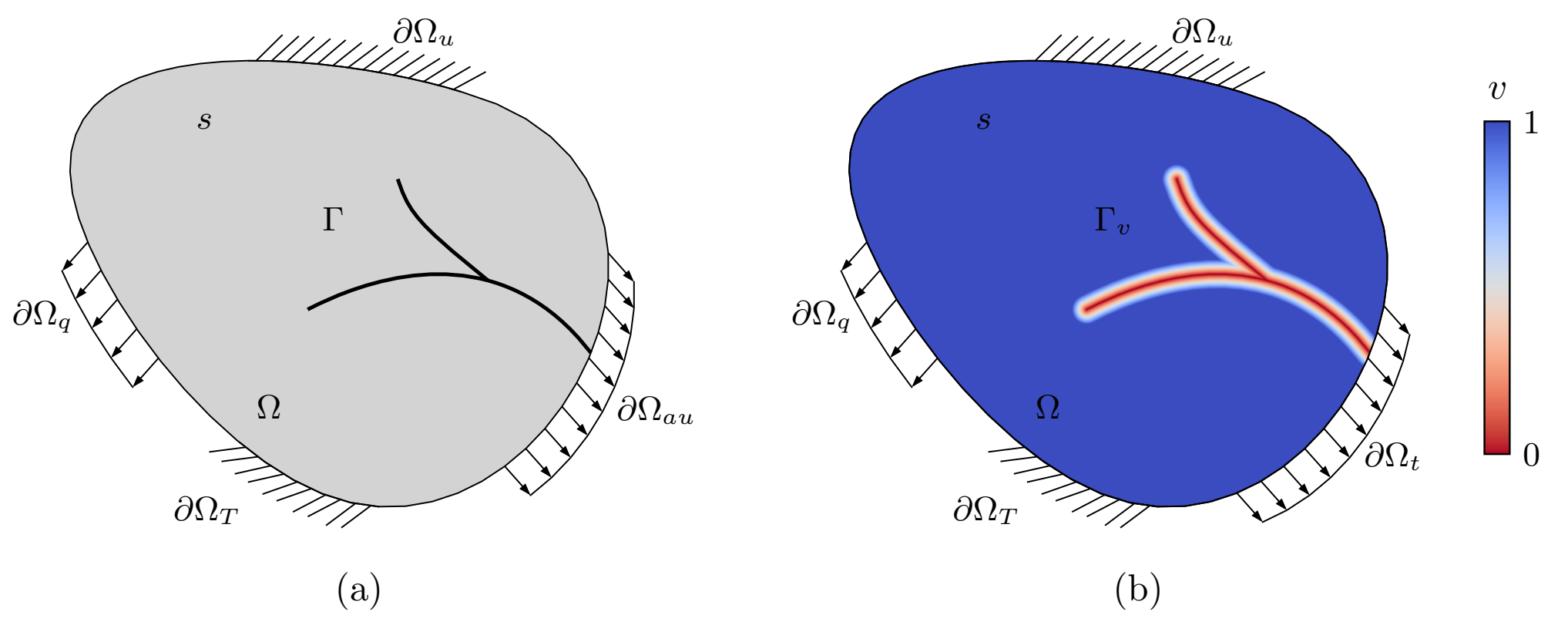

**Figure 1** Illustration of (a) the sharp crack model and (b) the general phase-field fracture model.

Consider a bounded domain $\Omega \subset \mathbb{R}^n$ with boundary $\partial\Omega$ and outward unit normal $\mathbf{n}$ subjected to thermo-mechanical loading. The state of the system from time $t \in (0, \mathrm{T}]$ is defined by three independent variables, the displacement field $\boldsymbol{u}(\mathbf{X}, t)$, the temperature field $T(\mathbf{X}, t)$, and the phase-field variable $v(\mathbf{X}, t)$, with $\mathbf{X}$ denoting the material point in the domain. The following subsections introduce each of the physical models that govern the fields, with reference to the schematic depicted in Figure 1(a).

## 2.1 The thermal problem formulation

In terms of the thermal loads, the boundary is partitioned into Dirichlet and Neumann parts. On the Dirichlet boundary, $\partial\Omega_T$, the temperature $\bar{T}$ is prescribed, whereas the heat flux $\bar{q}$ is prescribed on the Neumann portion of the boundary $\partial\Omega_q$.

In this work, $\rho$ is the density, $c_p$ the specific heat, $\boldsymbol{\kappa}$ the thermal conductivity tensor[2], and $s$ is the volumetric heat source. The heat equation is given by

$$\rho c_p \frac{\partial T}{\partial t} = -\nabla \cdot \mathbf{q} + s, \quad (\mathbf{X}, t) \in \Omega \times [0, \mathrm{T}], \tag{1}$$

[2] We note that, in general, the thermal properties can vary with position and temperature. These dependencies have been suppressed for the sake of conciseness.

with boundary conditions

$$T = \bar{T}, \quad (\mathbf{X}, t) \in \partial\Omega_T \times [0, \mathrm{T}], \tag{2a}$$

$$\mathbf{q} \cdot \mathbf{n} = \bar{q}, \quad (\mathbf{X}, t) \in \partial\Omega_q \times [0, \mathrm{T}], \tag{2b}$$

and initial conditions

$$T(\mathbf{X}, 0) = T_0(\mathbf{X}), \quad \mathbf{X} \in \Omega. \tag{3}$$

Finally, the heat flux density $\mathbf{q}$ is given by

$$\mathbf{q} = -\boldsymbol{\kappa} \cdot \nabla T. \tag{4}$$

In this work, the temperature field is coupled with the displacement field but not with the phase field. The thermal properties are assumed to be independent from the phase field, meaning the heat transfer across a regularized crack is undisturbed. For additional discussion and alternative formulations for coupling thermal and fracture effects in a regularized setting, see (Svolos et al. 2020).

## 2.2 The mechanical problem formulation

From time $t \in (0, \mathrm{T}]$, the position of a material point in the domain moves from $\mathbf{X}$ to $\mathbf{x}$:

$$\mathbf{x} = \mathbf{X} + \boldsymbol{u}(\mathbf{X}, t), \tag{5}$$

where $\boldsymbol{u}(\mathbf{X}, t)$ is the displacement field. We assume small deformations and deformation gradients, such that the deformation is described by the infinitesimal strain tensor

$$\mathbf{E}(\boldsymbol{u}) = \frac{1}{2}(\nabla \boldsymbol{u} + \nabla \boldsymbol{u}^T). \tag{6}$$

The mechanical part of the strain $\mathbf{E}_m$ is given by

$$\mathbf{E}_m(\boldsymbol{u}) = \mathbf{E}(\boldsymbol{u}) - \mathbf{E}_T, \tag{7}$$

where the thermal strain $\mathbf{E}_T$ is given by the coefficient of thermal expansion (CTE) $\alpha$ and the temperature difference between the current temperature $T$ and the reference temperature $\tilde{T}$:

$$\mathbf{E}_T = \alpha(T(\mathbf{X}, t) - \tilde{T})\mathbf{I}, \tag{8}$$

where $\mathbf{I}$ is the identity tensor.

The constitutive relation for the degraded Cauchy stress $\boldsymbol{\sigma}$ is

$$\boldsymbol{\sigma} = g(v)\frac{\partial \psi_{\text{ela}}}{\partial \mathbf{E}_m}(\mathbf{E}_m(\boldsymbol{u})), \tag{9}$$

with quadratic degradation function (Bourdin et al. 2000)

$$g(v) = v^2, \tag{10}$$

and the stored energy function

$$\psi_{\text{ela}}(\mathbf{E}_m(\boldsymbol{u})) = \frac{1}{2}\lambda(\operatorname{tr}\mathbf{E}_m(\boldsymbol{u}))^2 + \mu \mathbf{E}_m(\boldsymbol{u}) : \mathbf{E}_m(\boldsymbol{u}), \tag{11}$$

with $\mu > 0$ and $\lambda > -2/3\mu$ the Lamé constants.

The linear momentum balance is given by

$$\nabla \cdot \boldsymbol{\sigma} + \boldsymbol{b} = \rho \ddot{\boldsymbol{u}}, \quad (\mathbf{X}, t) \in \Omega \times [0, \mathrm{T}], \tag{12}$$

with boundary conditions

$$\boldsymbol{u} = \bar{\boldsymbol{u}}, \quad (\mathbf{X}, t) \in \partial\Omega_u \times [0, \mathrm{T}], \tag{13a}$$

$$\boldsymbol{\sigma} \cdot \mathbf{n} = \bar{\boldsymbol{\tau}}, \quad (\mathbf{X}, t) \in \partial\Omega_\tau \times [0, \mathrm{T}], \tag{13b}$$

with $\boldsymbol{b}$ denoting the body force, and $\bar{\boldsymbol{\tau}}$ the surface traction. The initial conditions are given by

$$\boldsymbol{u}(\mathbf{X}, 0) = \boldsymbol{u}_0, \quad \mathbf{X} \in \Omega, \tag{14a}$$

$$\dot{\boldsymbol{u}}(\mathbf{X}, 0) = \dot{\boldsymbol{u}}_0, \quad \mathbf{X} \in \Omega. \tag{14b}$$

## 2.3 The phase-field fracture model

The mechanical loading and/or the thermal loading give rise to deformation and potentially crack nucleation and propagation. The sharp crack $\Gamma$ is approximated by a diffuse crack $\Gamma_v$, as shown in Figure 1. It is described by the phase field

$$v = v(\mathbf{X}, t), \tag{15}$$

with $v = 0$ denoting fully fractured material and $v = 1$ intact material, as indicated in Figure 1(b).

The governing equation for the micro-force balance is given by:

$$\left.\begin{aligned} -g'(v)\psi_{\text{ela}} + \frac{3\delta G_c}{8\ell}(1 + 2\ell^2\Delta v) + c_e(\boldsymbol{u}, v) = 0 \quad & \text{if } \dot{v}(\mathbf{X}, t) < 0 \\ -g'(v)\psi_{\text{ela}} + \frac{3\delta G_c}{8\ell}(1 + 2\ell^2\Delta v) + c_e(\boldsymbol{u}, v) \geq 0, \quad & \text{if } \dot{v}(\mathbf{X}, t) \geq 0 \\ \dot{v}(\mathbf{X}, t) = 0, \quad & \text{if } v(\mathbf{X}, t) = 0 \end{aligned}\right\}, \quad (\mathbf{X}, t) \in \Omega \times [0, \mathrm{T}] \tag{16}$$

with boundary condition

$$\nabla v \cdot \mathbf{n} = 0, \quad (\mathbf{X}, t) \in \partial\Omega \times [0, \mathrm{T}], \tag{17}$$

and initial condition

$$v(\mathbf{X}, 0) = v_0(\mathbf{X}), \quad \mathbf{X} \in \Omega. \tag{18}$$

In the above, $G_c$ denotes the fracture toughness (also referred to as the critical fracture energy release rate), and $\ell$ denotes the regularization parameter (with units of length).

The model incorporates an arbitrary material strength surface by adding external microforces $c_e$ (Kumar, Ravi-Chandar, et al. 2018; Kumar et al. 2020; Kumar et al. 2022; Larsen et al. 2024). Importantly, the model does not belong to the class of classical variational phase-field models in which the evolution equations for the fields follow from a minimization of a global functional. Rather, the external microforces are brought into the model at the level of the governing equations, as opposed to the variational problem itself. As such, the external microforces only appear in the evolution equation for the phase field.

Physically, the external microforces can be viewed as the macroscopic manifestation of the presence of microscopic defects in the material. In terms of the governing equations (16), they act to enforce the requirement that under spatially uniform states of stress, the phase field $v$ does not decrease until the strength surface is violated. As noted in several previous works describing the complete model for purely mechanical problems (Kumar, Ravi-Chandar, et al. 2018; Kumar et al. 2020), the external microforces do dissipate energy. The effect of their dissipation could be included as an additional source term in the heat equation (1), but in this work we have assumed such contributions to be negligible.

The particular form of $c_e(\boldsymbol{u}, v)$ depends on the type of strength surface $\mathcal{F}(\boldsymbol{\sigma}) = 0$ employed. In this paper, we adopt the Drucker and Prager (1952) strength surface determined by

$$\hat{\mathcal{F}}(I_1, J_2) = \sqrt{J_2} + \frac{\sigma_{\text{ts}}}{\sqrt{3}(3\sigma_{\text{hs}} - \sigma_{\text{ts}})} I_1 - \frac{\sqrt{3}\sigma_{\text{hs}}\sigma_{\text{ts}}}{3\sigma_{\text{hs}} - \sigma_{\text{ts}}} = 0, \tag{19}$$

where $\sigma_{\text{hs}}$ and $\sigma_{\text{ts}}$ denote the hydrostatic and uniaxial tensile strength, respectively, and the $I_1$ and $J_2$ are invariants for the degraded Cauchy stress $\boldsymbol{\sigma}$:

$$I_1 = \operatorname{tr}(\boldsymbol{\sigma}) = \sigma_{11} + \sigma_{22} + \sigma_{33}, \tag{20}$$

$$J_2 = \frac{1}{2}\left(\boldsymbol{\sigma} - \frac{1}{3}\operatorname{tr}\boldsymbol{\sigma}\mathbf{I}\right) : \left(\boldsymbol{\sigma} - \frac{1}{3}\operatorname{tr}\boldsymbol{\sigma}\mathbf{I}\right), \tag{21}$$

To incorporate Equation (19) into the complete phase-field model, the external microforce $c_e$ takes the form of (Larsen et al. 2024; Kamarei et al. 2024):

$$c_e(I_1, J_2; \ell) = \beta_2\sqrt{J_2} + \beta_1 I_1 + v(1 - \operatorname{sgn}(I_1))\psi_{\text{ela}}, \tag{22}$$

with

$$\beta_1 = -\frac{1}{\sigma_{hs}(\mathbf{X})}\frac{\delta G_c}{8\ell} + \frac{2\psi_{hs}}{3\sigma_{hs}(\mathbf{X})}, \quad \beta_2 = -\frac{\sqrt{3}(3\sigma_{hs}(\mathbf{X}) - \sigma_{ts}(\mathbf{X}))}{\sigma_{hs}(\mathbf{X})\sigma_{ts}(\mathbf{X})}\frac{\delta G_c}{8\ell} - \frac{2\psi_{hs}}{\sqrt{3}\sigma_{hs}(\mathbf{X})} + \frac{2\sqrt{3}\psi_{ts}}{\sqrt{3}\sigma_{ts}(\mathbf{X})}, \tag{23}$$

and

$$\psi_{ts} = \frac{\sigma_{ts}^2(\mathbf{X})}{2E}, \quad \psi_{hs} = \frac{\sigma_{hs}^2(\mathbf{X})}{2K}, \tag{24}$$

where $E$ and $K$ denote the Young's modulus and bulk elastic modulus, respectively. Material strength in this work is generally considered to be spatially varied.

The strength surface (19) gives rise to a uniaxial compressive strength given by

$$\sigma_{cs} = \frac{2\sigma_{hs}\sigma_{ts}}{3\sigma_{hs} + \sigma_{ts}}. \tag{25}$$

The parameter $\delta$ is designed so that the model recovers Griffith-like crack propagation. It is given by (Kamarei et al. 2024)

$$\delta = \frac{\sigma_{ts} + (1 + 2\sqrt{3})\sigma_{hs}}{(8 + 3\sqrt{3})\sigma_{hs}}\frac{3G_c}{16\psi_{ts}\ell} + \frac{2}{5}. \tag{26}$$

The regularization length $\ell$ should be selected to be smaller than the smallest material length in the governing equations; see (Larsen et al. 2024) for additional details. A generally adopted upper bound is $\ell < \frac{3G_c E}{8\sigma_{ts}^2}$. For finite regularization lengths, the model gives rise to an effective strength surface given by

$$\hat{\mathcal{F}}^\ell(I_1, J_2) = \frac{J_2}{\mu} + \frac{I_1^2}{9K} - c_e(I_1, J_2; \ell) - \frac{3\delta G_c}{8\ell} = 0. \tag{27}$$

In the limit as $\ell \longrightarrow 0$, this approximate surface approaches the Drucker-Prager surface (19), see Figure 2.

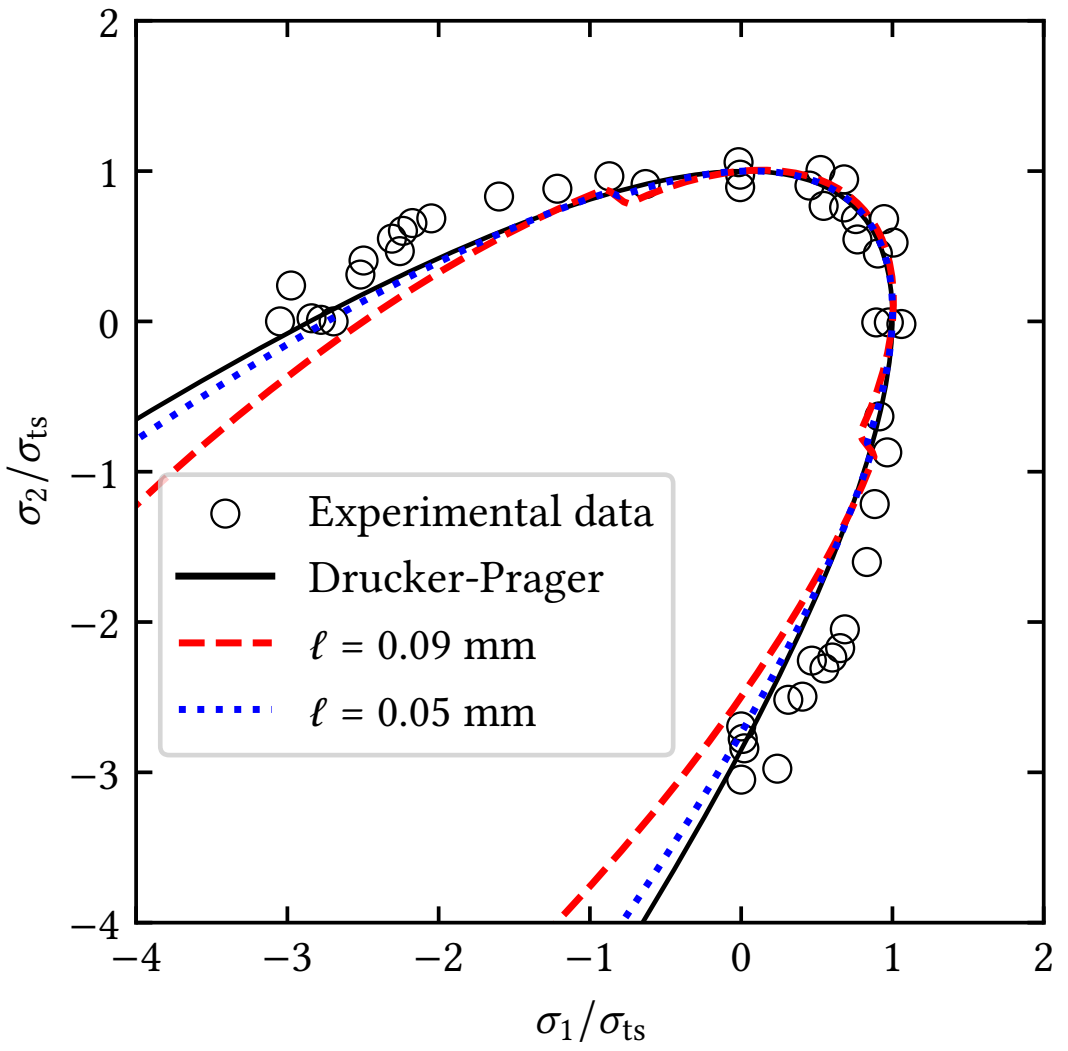

**Figure 2** Comparison between the experimental strength of graphite by Sato et al. (1987), the Drucker-Prager strength surface (19) defined by $\sigma_{ts}$ = 27 MPa and $\sigma_{hs}$ = 27.72 MPa, and the approximate strength surface (27) for two different regularization lengths. The approximate strength surface (dashed lines) asymptotes to the Drucker-Prager surface as the regularization length $\ell$ decreases.

For a complete list of the governing equations, see Box 1.

## 2.4 Spatial and temporal discretizations

To find approximations to the solutions for the system of equations presented in the previous subsections, we use standard finite elements in space and finite differences in time. Unless otherwise indicated, the displacement, temperature, and phase field are approximated using standard four-node quadrilateral elements for two-dimensional problems and eight-node hexahedral elements for three-dimensional problems.

Thermal:

$$
\begin{aligned}
&\rho c_p \frac{\partial T}{\partial t} = -\nabla \cdot \mathbf{q} + s, && (\mathbf{X}, t) \in \Omega \times [0, \mathrm{T}],\\
&q = -\kappa \cdot \nabla T, && (\mathbf{X}, t) \in \Omega \times [0, \mathrm{T}],\\
&T = \bar{T}, && (\mathbf{X}, t) \in \partial\Omega_T \times [0, \mathrm{T}],\\
&\mathbf{q} \cdot \mathbf{n} = \bar{q}, && (\mathbf{X}, t) \in \partial\Omega_q \times [0, \mathrm{T}],\\
&T(\mathbf{X}, 0) = T_0(\mathbf{X}), && \mathbf{X} \in \Omega,
\end{aligned}
$$

Mechanical:

$$
\begin{aligned}
&\nabla \cdot \left( v^2 \frac{\partial \psi_{\text{ela}}}{\partial \mathbf{E}_m} (\mathbf{E}_m(\boldsymbol{u})) \right) + \boldsymbol{b} = \rho \ddot{\boldsymbol{u}}, && (\mathbf{X}, t) \in \Omega \times [0, \mathrm{T}]\\
&\boldsymbol{u} = \bar{\boldsymbol{u}}, && (\mathbf{X}, t) \in \partial\Omega_u \times [0, \mathrm{T}],\\
&\left( v^2 \frac{\partial \psi_{\text{ela}}}{\partial \mathbf{E}_m} (\mathbf{E}_m(\boldsymbol{u})) \right) \cdot \mathbf{n} = \bar{\boldsymbol{\tau}}, && (\mathbf{X}, t) \in \partial\Omega_\tau \times [0, \mathrm{T}],\\
&\boldsymbol{u}(\mathbf{X}, 0) = \boldsymbol{u}_0(\mathbf{X}), && \mathbf{X} \in \Omega,\\
&\dot{\boldsymbol{u}}(\mathbf{X}, 0) = \dot{\boldsymbol{u}}_0(\mathbf{X}), && \mathbf{X} \in \Omega,
\end{aligned}
$$

Fracture:

$$
\begin{aligned}
&-g'(v)\psi_{\text{ela}} + \frac{3\delta G_c}{8\ell}(1 + 2\ell^2 \Delta v) + c_e(\boldsymbol{u}, v) = 0 && \text{if } \dot{v}(\mathbf{X}, t) < 0 \quad (\mathbf{X}, t) \in \Omega \times [0, \mathrm{T}],\\
&-g'(v)\psi_{\text{ela}} + \frac{3\delta G_c}{8\ell}(1 + 2\ell^2 \Delta v) + c_e(\boldsymbol{u}, v) \geq 0, && \text{if } \dot{v}(\mathbf{X}, t) \geq 0 \quad (\mathbf{X}, t) \in \Omega \times [0, \mathrm{T}],\\
&\dot{v}(\mathbf{X}, t) = 0, && \text{if } v(\mathbf{X}, t) = 0 \quad (\mathbf{X}, t) \in \Omega \times [0, \mathrm{T}],\\
&\nabla v \cdot \mathbf{n} = 0, && (\mathbf{X}, t) \in \partial\Omega \times [0, \mathrm{T}],\\
&v(\mathbf{X}, 0) = v_0(\mathbf{X}), && \mathbf{X} \in \Omega.
\end{aligned}
$$

**Box 1** Complete list of the governing equations.

In this work, the thermal subproblem is one-way coupled with the mechanical subproblem, while the mechanical and phase-field subproblems are fully coupled. At the beginning of each time step, the temperature field is updated and remains fixed. A staggered scheme (Hong and Luo 2017) is then used to iteratively find the optimal solution between the mechanical and phase-field subproblems. The irreversibility constraint on the phase-field parameter $v$ is imposed using a Primal-Dual Active Set strategy (Heister et al. 2015). In practice, the irreversibility is only enforced once $v$ drops below a threshold, in other words when $v \leq v_{\text{th}}$. In this work, $v_{\text{th}} = 0.05$.

For the temporal discretization, the mechanical and thermal subproblems are treated using different integrators. The thermal solve is discretized with a first-order implicit method. For the mechanical subproblem, temporal discretization is only required when inertial effects are included. In these cases, the HHT-$\alpha$ scheme (Hilber et al. 1977) is adopted to discretize the elastodynamics subproblem. In particular, the HHT-$\alpha$ scheme is employed with the parameters $\alpha^{\text{hht}} = -0.3$, $\beta^{\text{hht}} = 0.4225$, and $\gamma^{\text{hht}} = 0.8$, for which the method is second-order accurate and unconditionally stable. For the thermal subproblem, a first-order implicit Euler scheme is used. While a second-order scheme could certainly be used to obtain consistency with displacement update when inertial effects are included, it was not found to be needed to obtain sufficiently accurate results for the problems considered here.

With the complete phase-field fracture model, it has been established that the discrete fracture toughness is influenced by the mesh size. To offset this effect, a mesh-dependent correction is introduced to the expression for $\delta$:

$$
\delta = \left(1 + \frac{3\text{h}}{8\ell}\right)^{-2} \frac{\sigma_{\text{ts}} + (1 + 2\sqrt{3})\sigma_{\text{hs}}}{(8 + 3\sqrt{3})\sigma_{\text{hs}}} \frac{3G_c}{16\psi_{\text{ts}}\ell} + \left(1 + \frac{3\text{h}}{8\ell}\right)^{-1} \frac{2}{5}. \tag{28}
$$

This correction was derived for linear triangular and tetrahedral elements. The mesh-dependent correction function particularly calibrated for quadrilateral/hexahedral-elements is

$$\delta = \left(1 + \frac{3\mathrm{h}}{8\ell}\right)^{-2} \frac{0.2201\sigma_{\mathrm{hs}} + 0.1295\sigma_{\mathrm{ts}}}{\sigma_{\mathrm{hs}}} \frac{3G_c}{16\psi_{\mathrm{ts}}\ell} + \left(1 + \frac{3\mathrm{h}}{8\ell}\right)^{-1} 0.6297. \tag{29}$$

While both functions ultimately yield the same effective fracture toughness, each converges more rapidly as $\ell$ decreases on its designated element types.

In all cases, results were obtained via implementation in RACCOON (Torres et al. 2026)[3]. The particular phase-field implementation employed in this work has been extensively verified and validated against the nine circles of elastic brittle fracture, as described in Kamarei et al. (2026).

# 3 Numerical Examples

To assess the capabilities of the complete phase-field fracture model in coupled thermo-mechanical settings, we consider three representative problems that span a range of fracture nucleation scenarios: propagation from a large pre-existing crack, nucleation under nearly uniform stresses, and intermediate cases combining localized and distributed stress fields. Together, these examples cover a broad spectrum of loading conditions, specimen geometries, and stress states, allowing us to evaluate the model's performance in predicting both the onset and evolution of fracture. In the following subsections, we provide the experimental configurations, the modeling assumptions and discretization details for each simulation result, along with comparisons between simulated and observed fracture patterns. These case studies not only demonstrate the model's versatility but also highlight the importance of incorporating material strength as an independent property in brittle fracture modeling. In all cases, the results shown were verified to be both spatially and temporally converged.

## 3.1 Progressive quenching of glass plates

In this section, we examine the ability of the complete model to capture crack evolution patterns in glass plates subjected to sudden but carefully controlled cooling. Since it was first proposed by Yuse and Sano (1993), this problem has been extensively studied, both experimentally (Ronsin et al. 1995; Yang and Ravi-Chandar 2001; Yoneyama et al. 2008) and numerically. Simulations of the experiment have been performed using both sharp (Bahr et al. 1995; Sumi and Mu 2000; Chiaramonte et al. 2020) and regularized models (Corson et al. 2009; Kilic and Madenci 2009; Menouillard and Belytschko 2011; Xu et al. 2018; Pan et al. 2024), as well as peridynamics (Xu et al. 2018). These simulations have reproduced the crack patterns under a variety of conditions, from relatively mild thermal shocks to higher thermal loads, to a limited degree. The problem is interesting because the conventional understanding is that it is explained by crack propagation as governed by energetics. However, in what follows, we present new results that illustrate how strength is an important factor that helps explain experimental observations.

Although there is some variation in aspects of the geometry of the glass plate, the loading condition mostly follows the standard depicted in Figure 3(a). In the experiment, a long thin glass specimen of width $2b$ with an initial crack at its leading edge is moved at a constant velocity $V$. It is first passed through an oven and heated to a desired uniform temperature $T_H$. Soon after the leading edge exits the oven, it moves into a cold bath of temperature $T_C$. As the plate keeps moving, fresh hot material is submerged into the bath, mimicking a quenching process. The transition from the warm oven to the cold bath gives rise to a sudden temperature gradient that induces a thermo-mechanical stress field.

The evolution of the large pre-existing crack at the leading edge of the plate has been experimentally observed to be governed by several factors. These include, for example, the geometry of the plate, its material properties, the applied velocity, and the gap between the oven and the bath. Roughly speaking, as the magnitude of the forcing is increased, a transition in the crack morphology has been observed. For small forcings, the initial crack simply propagates in a straight line from its initial orientation, as shown in the top image in Figure 3(b). As the forcing is increased, this morphology transitions to an oscillatory pattern as shown in the middle panel of Figure 3(b). Finally, for sufficiently large forcings, a snap back in the crack evolution can be observed as well as much more irregular propagation patterns. An example is the pattern shown in the bottom panel of Figure 3(b).

[3] https://hugary1995.github.io/raccoon/index.html, an open-source, massively parallel finite element code built upon the MOOSE framework (Giudicelli et al. 2024; Gaston et al. 2015; Lindsay et al. 2021)

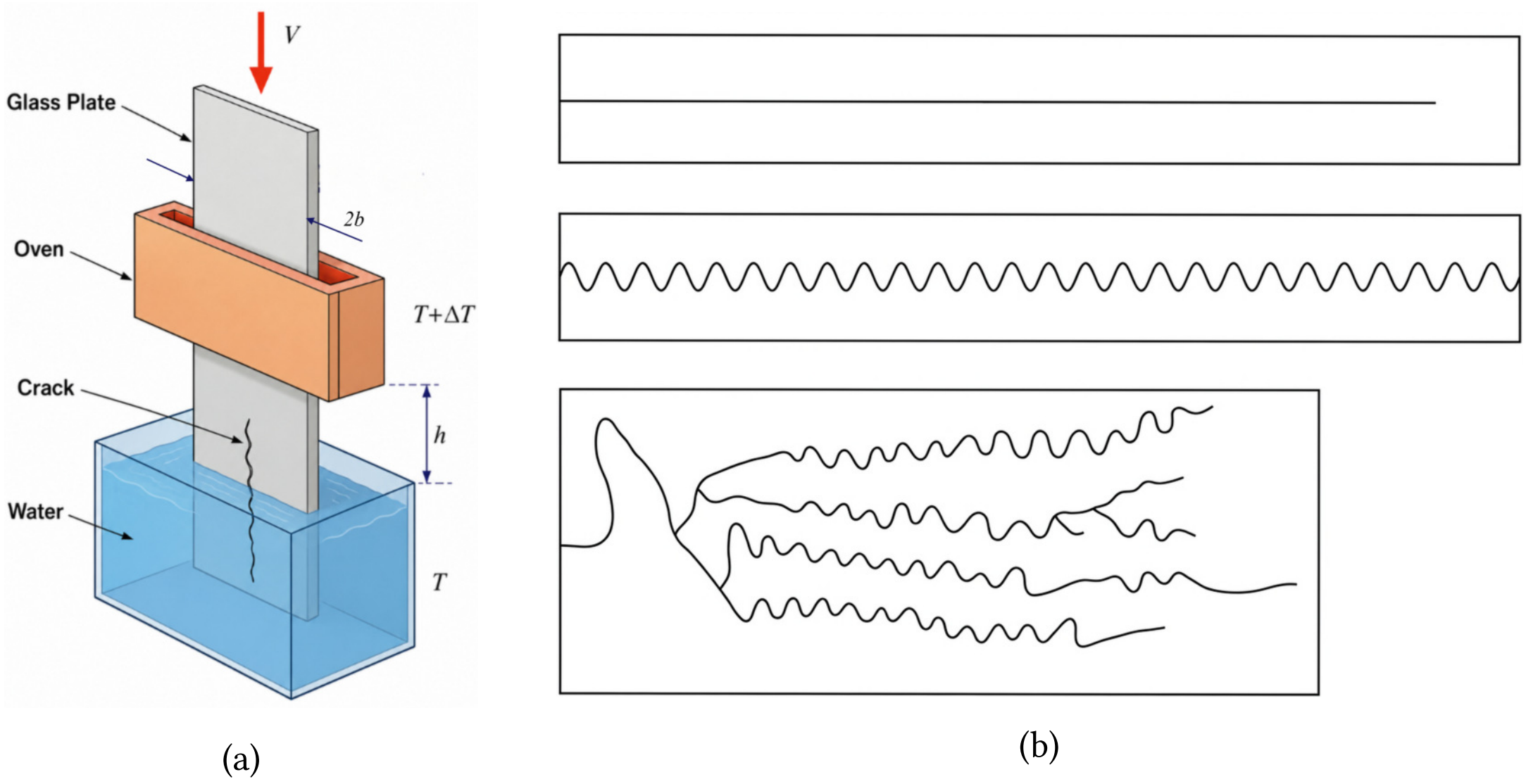

**Figure 3** Experimental setup and observed crack patterns for progressive quenching of glass plates. (a) Schematic of the experimental setup for the progressive quenching of glass plates and (b) a collection of experimental crack morphology photos from (Yuse and Sano 1997; Sumi and Mu 2000), providing one example of straight crack propagation, one example of regular oscillatory propagation, and one example of chaotic propagation.

In this paper, we focus on configurations of the experiment with plate thickness $e$, oven-bath gap $h$, and thermal diffusion length $D/V$ satisfying $e < D/V < h$, where $D$ is the plate's thermal diffusion coefficient. Such configurations ensure that the temperature field is homogeneous across the thickness, and the temperature field in the velocity direction is a function of the ratio $D/V$. For further discussion of the influence of various aspects of the experimental setup on the resulting thermal fields, see (Ronsin et al. 1995).

We begin our analysis of this configuration by developing an analytical expression for the temperature field in the plate. We assume that the diffusion length is much smaller than the size of the gap between the bath and the oven, such that any cooling that occurs between them can be neglected. Along the length of the plate, the temperature field can be divided into three sections as shown in Figure 4. Roughly speaking, these consist of the portion of the plate fully immersed in the cold bath, the portion fully in the oven, and the region in between over which a gradient in the temperature field develops.

For concreteness, we define a set of coordinate axes $\mathbf{e}_x, \mathbf{e}_y$ to be oriented with the direction of the plate velocity and its height, as shown in Figure 4. The position of the cold immersion front is denoted by $x_C$, and so the temperature at points where $x \leq x_C$ is assumed to match the bath temperature $T_C = 23\,°\mathrm{C}$. The temperature evolution between the cold immersion front and the uniformly heated region can be described by the advection-diffusion equation

$$V\frac{dT}{dx} = D\frac{d^2T}{dx^2}. \tag{30}$$

The analytical solution to this one-dimensional thermal advection problem with a moving cold front is given by

$$T(x,t) = T_H - (T_H - T_C)\exp\left(\frac{-V}{D}(x - V \times t)\right), \tag{31}$$

which is depicted in purple in Figure 4. In essence, as the ratio $V/D$ is increased, the gradient in the temperature field becomes steeper near the immersion front. In what follows, all of the results were generated assuming this functional form of the temperature field.

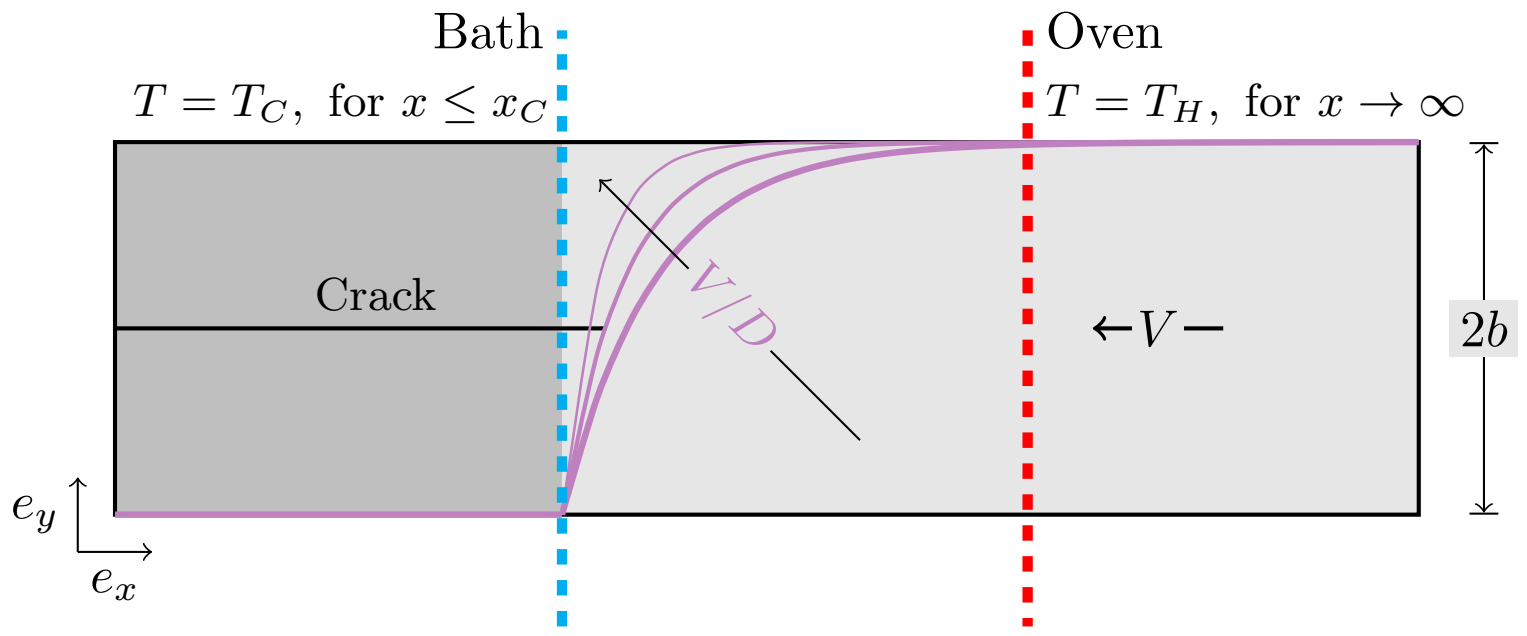

**Figure 4** Schematic of the simulated geometry. The evolution of the temperature field as a function of $V/D$ between the hot zone and cold front are as illustrated by the purple curves.

**Table 1** Glass material properties (Yang and Ravi-Chandar 2001).

| | symbol | unit | R.T. |
|---|---|---|---|
| Young's modulus | E | GPa | 70 |
| Poisson's ratio | $\nu$ | | 0.22 |
| Density | $\rho$ | $10^3$ kg m$^{-3}$ | 2.5 |
| Critical stress intensity factor | $K_c$ | MPa$\sqrt{\text{m}}$ | 0.76 |
| Fracture toughness | $G_c$ | N m$^{-1}$ | 8.25 |
| Thermal conductivity | $\kappa$ | W m$^{-1}$ K$^{-1}$ | 1.08 |
| Thermal diffusivity | $D$ | $10^{-6}$ m$^2$ s$^{-1}$ | 0.47 |
| Specific heat | $c_p$ | J kg$^{-1}$ K$^{-1}$ | 921 |
| CTE | $\alpha$ | $10^{-6}$ K$^{-1}$ | 7.7 |
| Tensile strength | $\sigma_{\text{ts}}$ | MPa | 80 |
| Compressive strength | $\sigma_{cs}$ | | $20\sigma_{\text{ts}}$ |

The material properties used for the model are provided in Table 1. Simulations were conducted for a plate with half width $b = 8$ mm and an initial crack length of 5 mm. The horizontal dimension of the plate was taken to be 30 mm, which was found to be sufficiently long to permit crack patterns to develop. The regularization length $\ell$ was set to 0.03 mm and an adaptively refined mesh with the finest mesh size of $h = \ell/5$ was used. To provide a small, symmetry-breaking perturbation, the initial crack is positioned with a uniform offset of +0.25 mm in the $y$-direction. As crack propagation occurs at speeds that are slow compared to the bulk wave speeds of the material, inertial effects were neglected for these simulations, and the displacement sub-problem was obtained as a quasi-static result. To fix the rigid body modes, displacement boundary conditions were prescribed as follows: the horizontal displacement along the entire far right edge of the domain was set to zero, and the vertical displacement of the center point along the same edge was also set to zero.

We begin by presenting qualitative reproductions of the observed crack pattern trends. In Figure 5, a series of crack patterns obtained using $V = 0.4$ mm/s and selected temperature jumps $\Delta T$ are shown. For $\Delta T < 47\,^\circ$C, no crack propagation is observed. For the range $67\,^\circ\text{C} \leq \Delta T \leq 127\,^\circ$C, cracks propagate along a purely straight path. As the temperature jump increases to 147–167 °C, the cracks follow a regular oscillatory trajectory. With further increases in $\Delta T$, branching occurs. At $\Delta T = 307\,^\circ$C, the initial crack first undergoes a snap-back, then continues to propagate forward in an irregular wavy pattern.

As the temperature jump increases, the energy to be released per unit crack advance in the quenching direction also increases. Consequently, crack paths that dissipate more energy by creating a longer crack trajectory for the same projected advancement become increasingly favorable. Larger temperature jumps also tend to give rise to increased stresses, thus creating more favorable conditions for additional crack nucleation. These considerations provide physical insights into the observed transitions between straight growth, to growth with regular oscillations, to crack growth with branching, and finally to snap-back or irregular wavy patterns at the highest thermal loads.

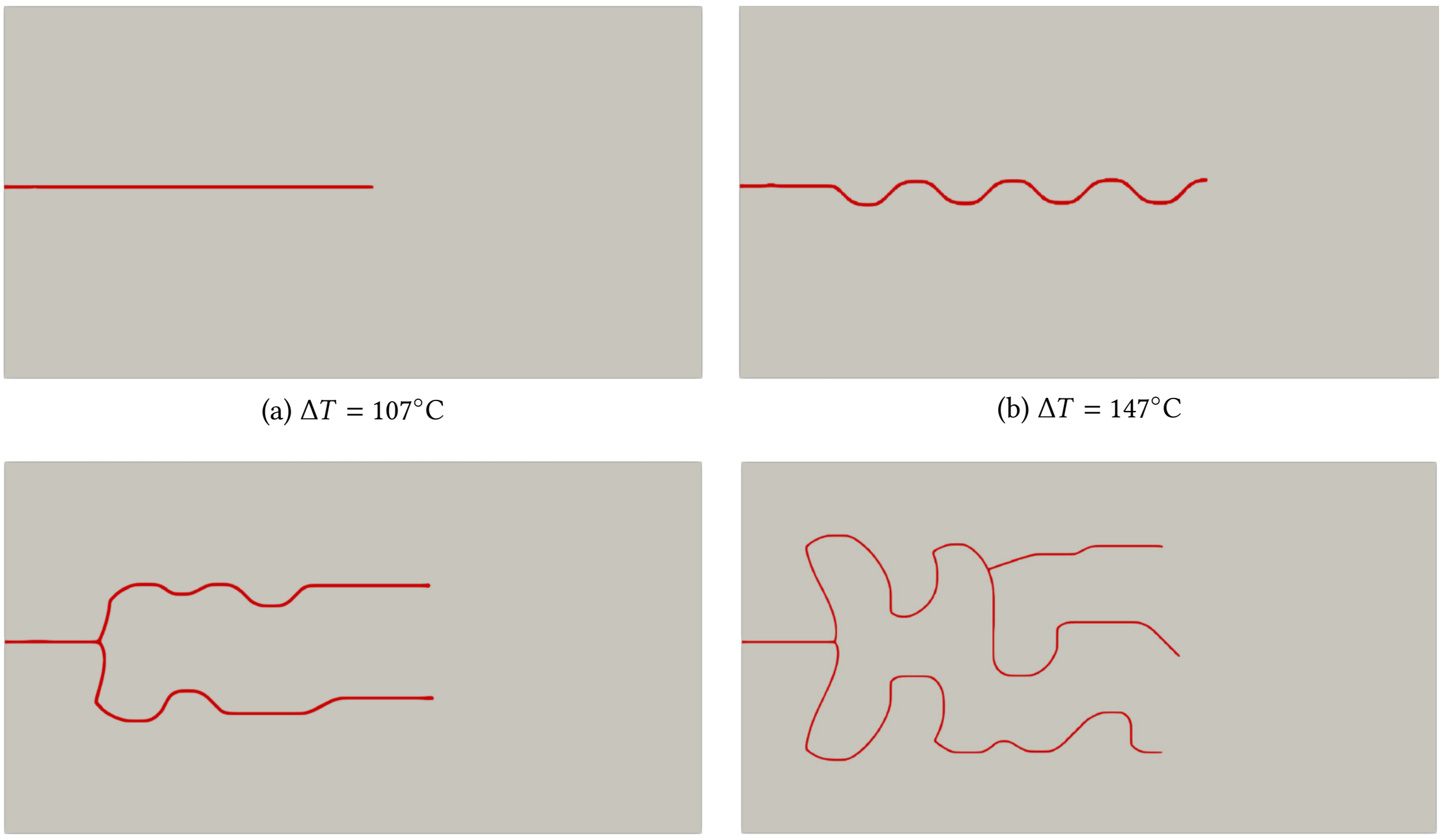

(a) $\Delta T = 107°\mathrm{C}$ (b) $\Delta T = 147°\mathrm{C}$

(c) $\Delta T = 187°\mathrm{C}$ (d) $\Delta T = 307°\mathrm{C}$

**Figure 5** Representative crack patterns extracted from simulations using the complete phase-field model with $V = 0.4$ mm/s and selected temperature jumps $\Delta T$. (a) straight crack propagation, (b) regular oscillatory propagation, (c) branching followed by oscillatory crack propagation that gradually transitions to straight propagation, (d) snap-back and branching, followed by chaotic oscillatory propagation and subsequent branching.

Based on multiple experiments carried out under different configurations, two non-dimensional quantities were proposed to delineate the various types of fracture responses (Corson et al. 2009; Chiaramonte et al. 2020). The first is the Peclet Number, which is the ratio of advective to diffusive effects:

$$P = \frac{bV}{D}. \tag{32}$$

The scope of this study is for the advection dominated regime, in which $P > 1$. The second dimensionless quantity is the ratio of the thermal shock to the fracture resistance:

$$R = \frac{\alpha E \Delta T \sqrt{b}}{K_c}, \tag{33}$$

with $\alpha$ the coefficient of thermal expansion, $E$ Young's modulus, and $\Delta T$ the temperature jump. In the above, $K_c$ refers to the critical stress intensity factor, which under plane-stress conditions is related to the fracture toughness $G_c$ by:

$$G_c = \frac{K_c^2}{E}. \tag{34}$$

In order to explore the formation and evolution of crack patterns, several crack paths were computed on a regular grid in the phase space $(P, R)$, as shown in Figure 6. The phase space is classified into five regions, each corresponding to a distinct fracture response. The boundaries between these regions are defined by transitions identified through experimental observations or previous simulation results that can be found in the literature (see caption). The simulation results obtained using the complete model are indicated by markers. Although computed on a relatively sparse grid in phase space, the transitions between marker types closely align with the borders established by previous studies.

The propagation of a long initial crack is typically regarded as an energetically-dominated process, wherein the Griffith-like competition between the potential energy and the fracture toughness governs the response. However, with the complete model, we now show that material strength is also influential and can even determine the final crack pattern. In Figure 7, we present two sets of snapshots of crack evolution obtained with different material strengths: with strengths as provided in Table 1, and a comparison example with strengths weakened to 25 %. In the first frame, both cases propagate the existing crack downward, forming a curved corner in the upper half of the specimen. In the second frame, the weaker material develops a stress state at the curved corner that exceeds its strength, leading to the nucleation of a second branch, ultimately forming a tuning-fork shaped pattern. By contrast, the stronger material withstands the loading at the curved corner and the crack propagates along a single oscillatory path. These results make it clear that, even for a problem involving the propagation of a large crack, it is important to use an accurate representation for the material strength. Interestingly, a similar change in crack patterns for this class of problems was reported in (Xu et al. 2018), in which results were obtained using peridynamics. In that work, changes were reported due to a change in the horizon length.

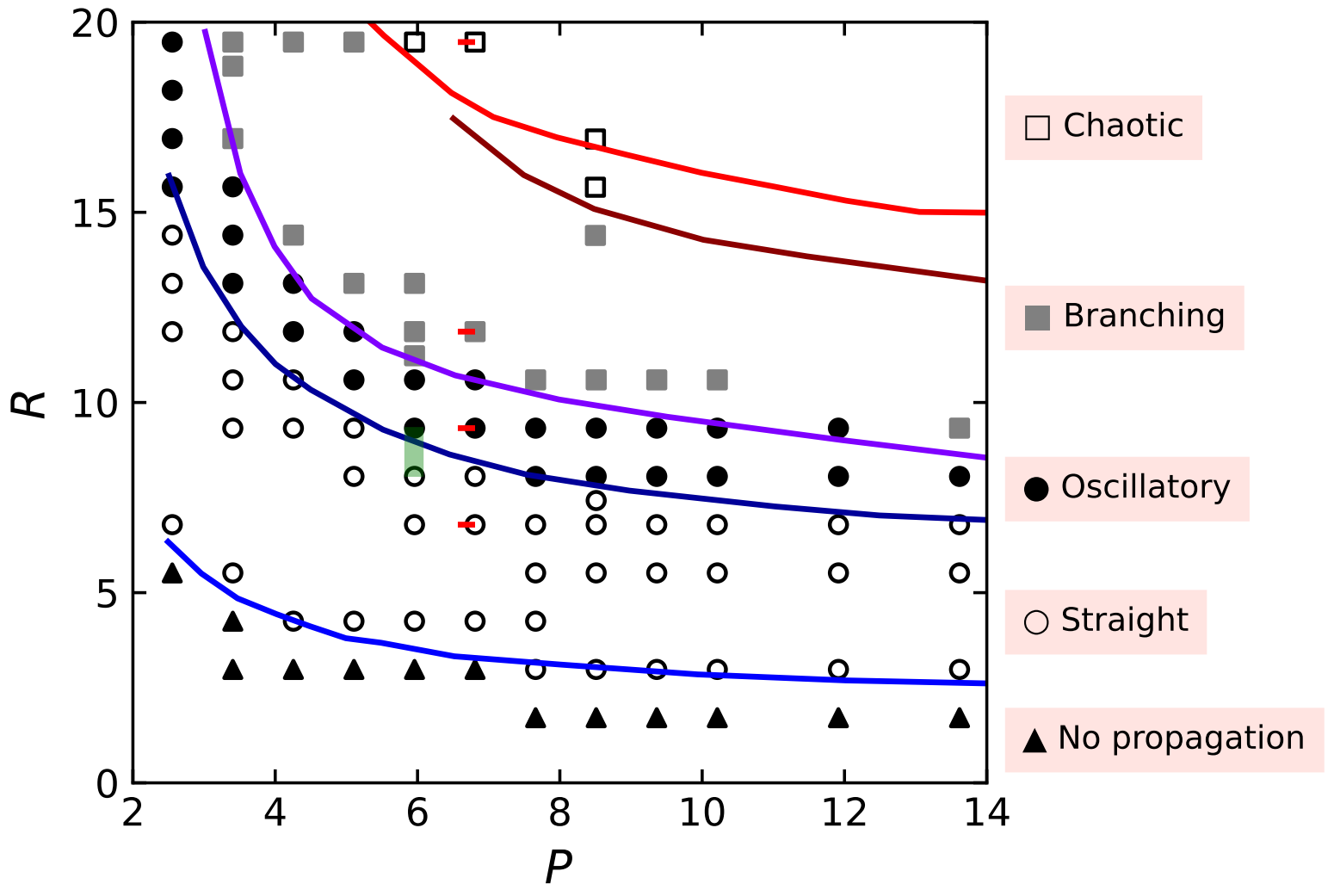

**Figure 6** Phase diagram of fracture patterns as delineated by the dimensionless parameters $P$ and $R$. The curves are crack pattern borders aggregated from a collection of experimental observations and previous simulation results. From the bottom to the top, each curve was extracted from (Ronsin et al. 1995; Ronsin and Perrin 1997; Chiaramonte et al. 2020; Chiaramonte et al. 2020; Yuse and Sano 1997). See (Chiaramonte et al. 2020) for a detailed explanation on these curves. The markers are the phase-field modeling results obtained using the complete model. The red curve represents the phase boundary extracted from the experiments of Yuse and Sano (1997), while the dark red curve represents the corresponding boundary obtained from the model of Chiaramonte et al. (2020). The markers with a left red tick each corresponds to the crack pattern displayed in Figure 5. The small area shaded in green at $P = 6$ and $8 \leq R < 9$ corresponds to the results in Figure 8.

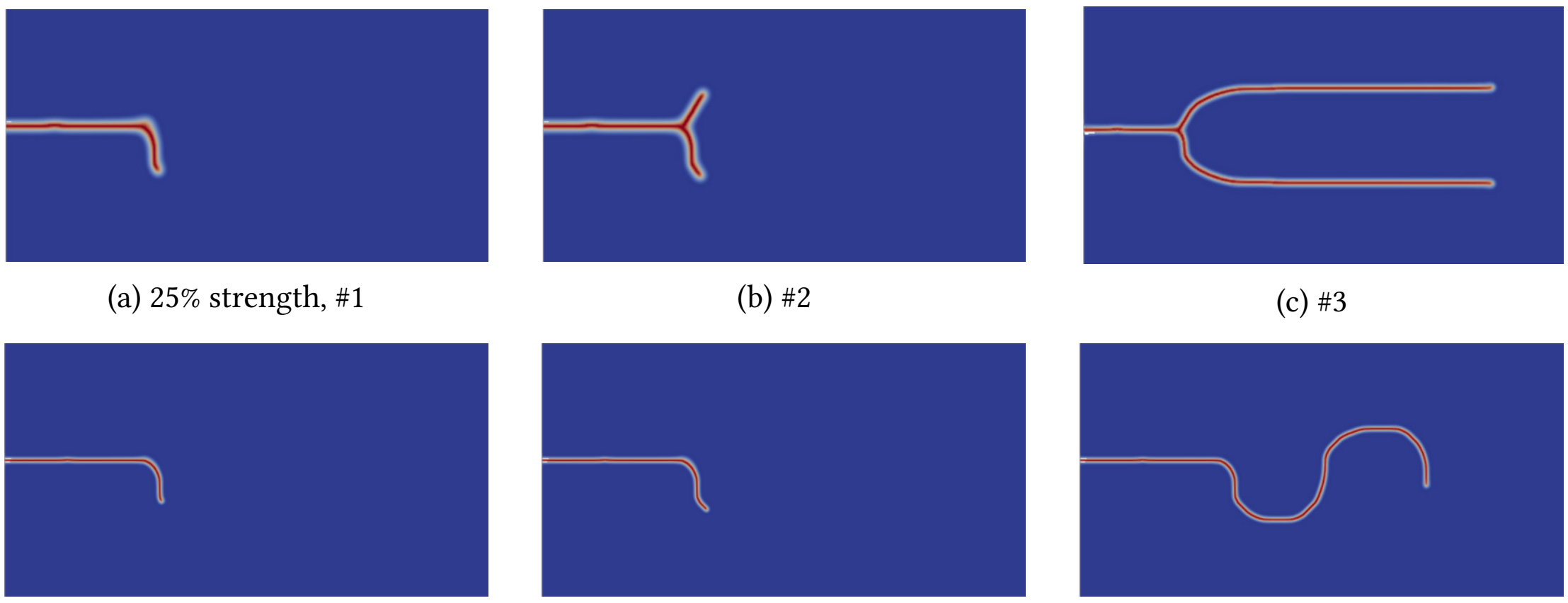

(a) 25% strength, #1 (b) #2 (c) #3

(d) 100% strength, #1 (e) #2 (f) #3

**Figure 7** Snapshots of crack propagation under the same temperature jump, but with different strength surfaces. (a)-(c) propagation pattern for a material with strength surface scaled to 25 % with respect to the origin. (d)-(f) from a sample with the original strength of glass, as specified in Table 1.

Of course, the material strength was not an input to the model (not even the uniaxial tensile strength), and so any sensitivity to the strength simply wasn't examined in (Xu et al. 2018).

Along these same lines, we note the comparison of our results to those of Chiaramonte et al. (2020) vs. the results reported in (Yuse and Sano 1997). The red and dark red curves in Figure 6 delineate regions of the phase diagram between branching (below) and chaotic (above). Chiaramonte et al. (2020) attributed the difference between their model-based results and the experimental observations to the use of a heuristic for crack branching, but our analysis indicates that the differences may be due to actual differences in the strengths of the various glasses. In other words, the experimental results shown in the phase diagram stem from studies on several different types of glasses, but the quantity $R$ as defined in Equation (33) does not contain any information about the strength of the material, only its toughness. These observations obviously call into question the use of the particular dimensionless quantities employed by previous researchers to construct the phase diagram.

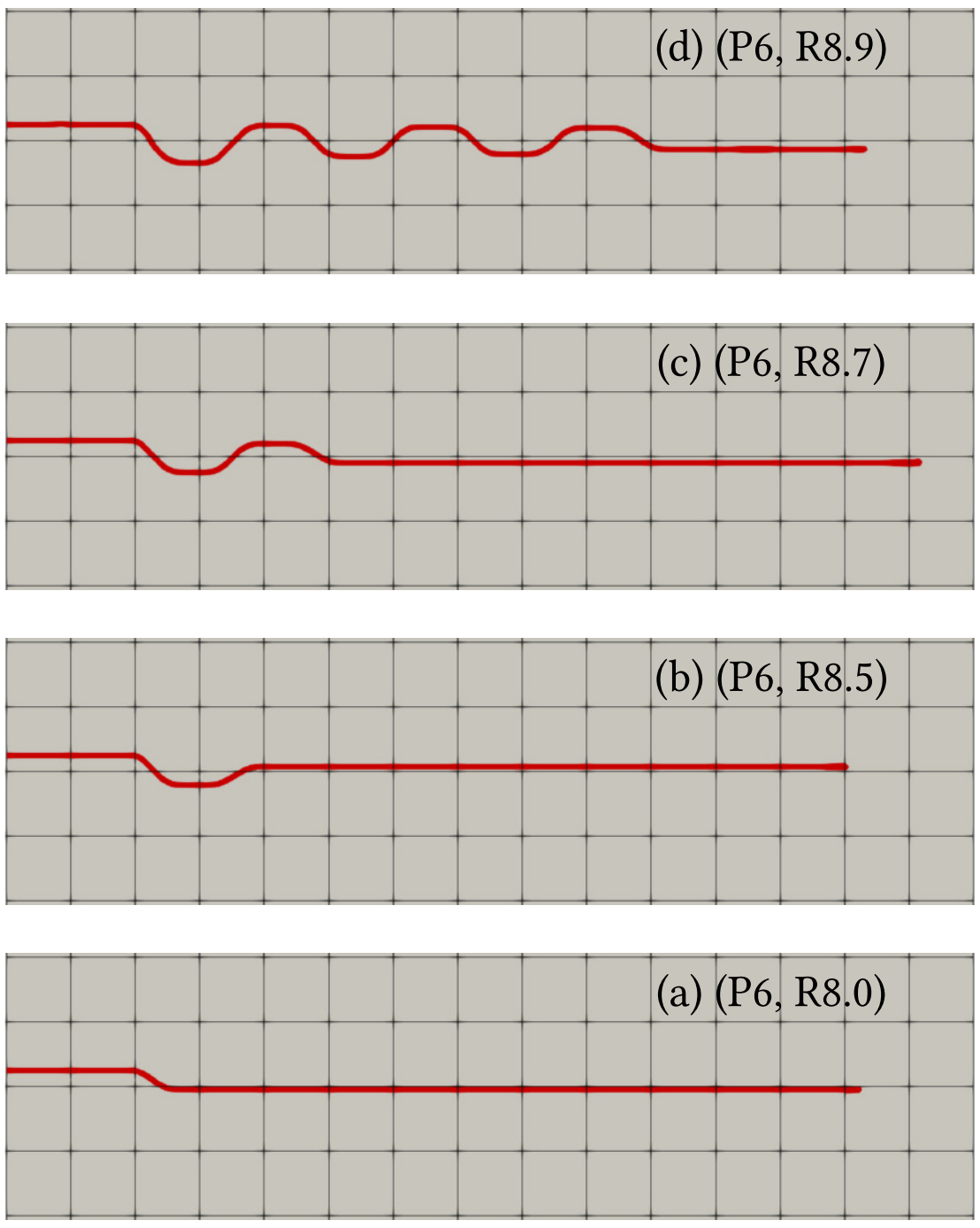

**Figure 8** Examples of crack path oscillation with decaying amplitude. (a)-(d) phase-field model results obtained from corresponding $(P, R)$ values that illustrate the trend. The superimposed reference grid has dimension $0.25b \times 0.25b$. The finite element mesh for each result is much finer and not shown.

At the phase border between straight and oscillatory fracture patterns (Figure 6), a gradual transition zone exists. Figure 8 shows a sequence of crack patterns for a fixed value of the Peclet Number $P$, corresponding to increasing values (from the bottom to the top) of the fracture driver number $R$. For the smaller values of $R$ (Figures 8(a) to 8(c)), the crack exhibits a few oscillations with decreasing amplitude, before returning to the centerline. As $R$ increases (Figure 8(d)), the number of oscillations increases. Beyond this transition zone, the path oscillation is steady, but the form of the oscillation changes as described below.

Figure 9 presents results corresponding to three representative cases that are fully within the oscillatory zone of the phase diagram. The plots contain information regarding the crack tip positions, the fracture patterns, and the crack tip velocities. They are presented in a style that is analogous to those provided in Figure 15 of (Chiaramonte et al. 2020), to facilitate comparisons. To obtain the results using the phase-field model, the crack tip position was identified based on the $v = 0.05$ contour. The crack tip velocity was then computed based on the total distance traveled over a window of two time steps.

The crack path obtained for $P = 6.8$ and $R = 9.3$, Figure 9(a), exhibits a sinusoidal wave pattern, with the first half and the second half of each period being symmetric and smooth changes in propagation direction. The crack tip velocity can be seen to remain relatively constant and comparable to the immersion speed.

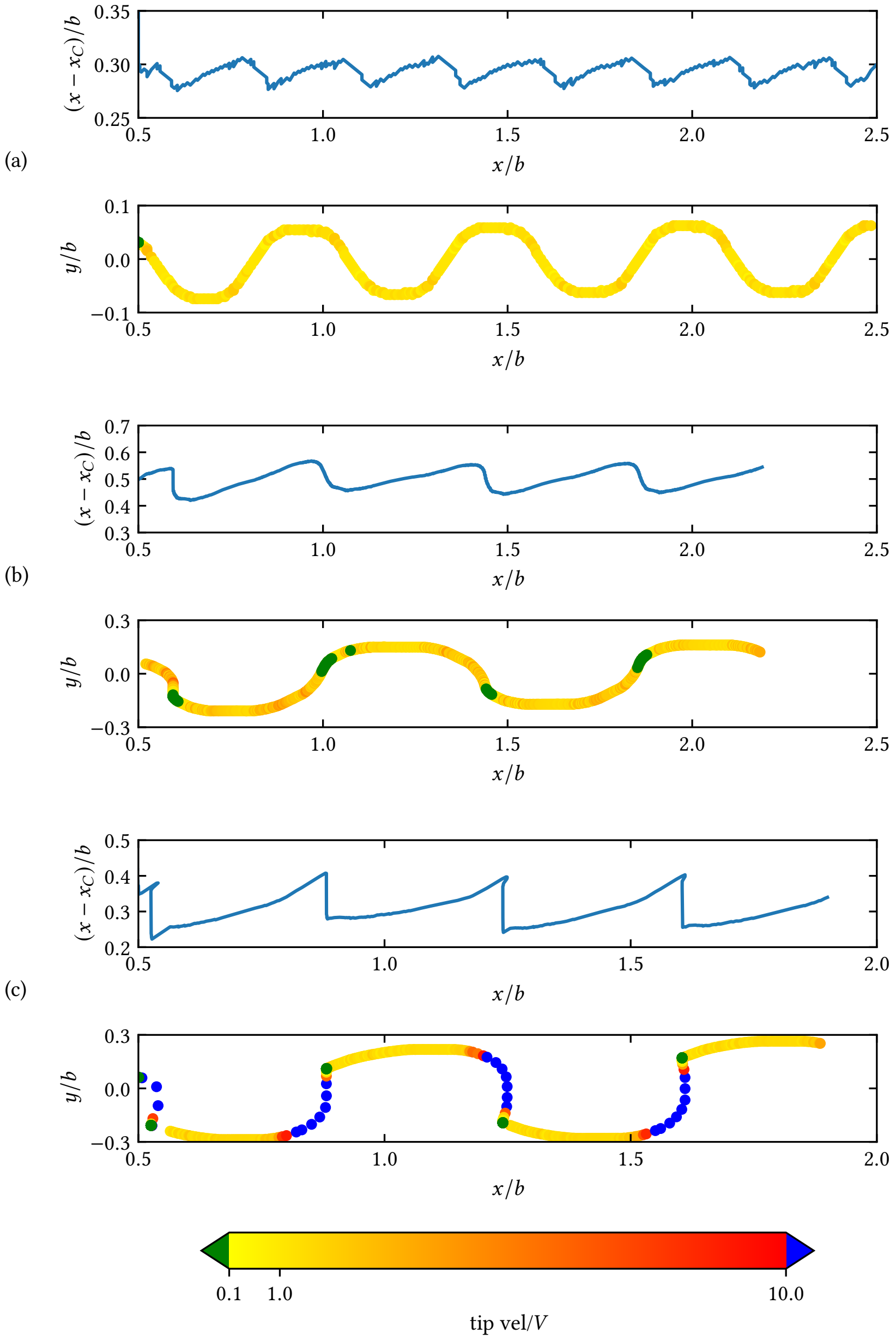

**Figure 9** Examples of crack-tip location and velocity profiles obtained at (a) ($P = 6.8, R = 9.3$), (b) ($P = 3.4, R = 13.1$), and (c) ($P = 6.8, R = 10.6$). For each case, the bottom plot shows a sequence of crack-tip locations, with each point colored according to the instantaneous crack-tip velocity (normalized by the immersion speed). The top plot presents the corresponding crack-tip position as a function of distance from the immersion front. Each of the three profiles represents a typical behavior previously reported in (Chiaramonte et al. 2020).

By contrast, the crack pattern obtained for $P = 3.4$ and $R = 13.1$, Figure 9(b), exhibits a crack tip velocity that varies in time, with intermittent stalling within a period. After a stall, the crack propagates first along the $x$ direction, then smoothly turns into the $y$ direction with a slightly increasing velocity, forming a flat-to-circular path. At the end of a short segment almost along the $y$ direction, it stalls again. The next semi-period repeats the behavior but turns in the opposite direction. The stall of the crack tip allows the immersion front to catch up and provide sufficient energy and an appropriate stress state in the vicinity to continue propagation.

Finally, the crack pattern obtained for $P = 6.8$ and $R = 10.6$, Figure 9(c), corresponds to crack tip velocities that vary over an order of magnitude. Similar to the previous example, after a stall point the crack first propagates almost along the $x$ direction at a speed close to the immersion speed. Its velocity then increases considerably as it travels along a curved path. The curve ends with a segment nearly perpendicular to the immersion direction, with some periods showing a slight snap-back. Each "half-period" ends at a stall point, from which the next "half-period" begins with a sharp turn that is nearly 90°.

### 3.1.1 Sensitivity to mesh and regularization length

As mentioned previously and illustrated in many previous works, the complete phase-field fracture model provides simulation results that are insensitive to the choice of mesh type or the value of $\ell$, provided the latter is taken to be sufficiently small. Results obtained using two different finite element meshes (of comparable resolution) constructed using either bilinear quadrilateral elements (quad4) or linear triangular elements (tri3) are provided in Figure 10(a). There are of course slight differences, but a clear influence of the mesh on the fracture patterns, beyond what is evident from the fact that the phase field $v$ is interpolated by the mesh, is not apparent. For all practical purposes, the results are nearly indistinguishable.

Results obtained using two different regularization lengths, one 50 % smaller than the other, are provided in Figure 10(b). For the smaller regularization length, the mesh spacing is also reduced by a factor of 2. Although some minor differences appear in the periodicity of the fracture pattern at later stages, the amplitude of the oscillations is indistinguishable. This is to be contrasted with comparable results for a similar problem obtained with peridynamics, in which a halving of the horizon length results in a dramatic change in the amplitude of the oscillations, see (Xu et al. 2018, Figure 13).

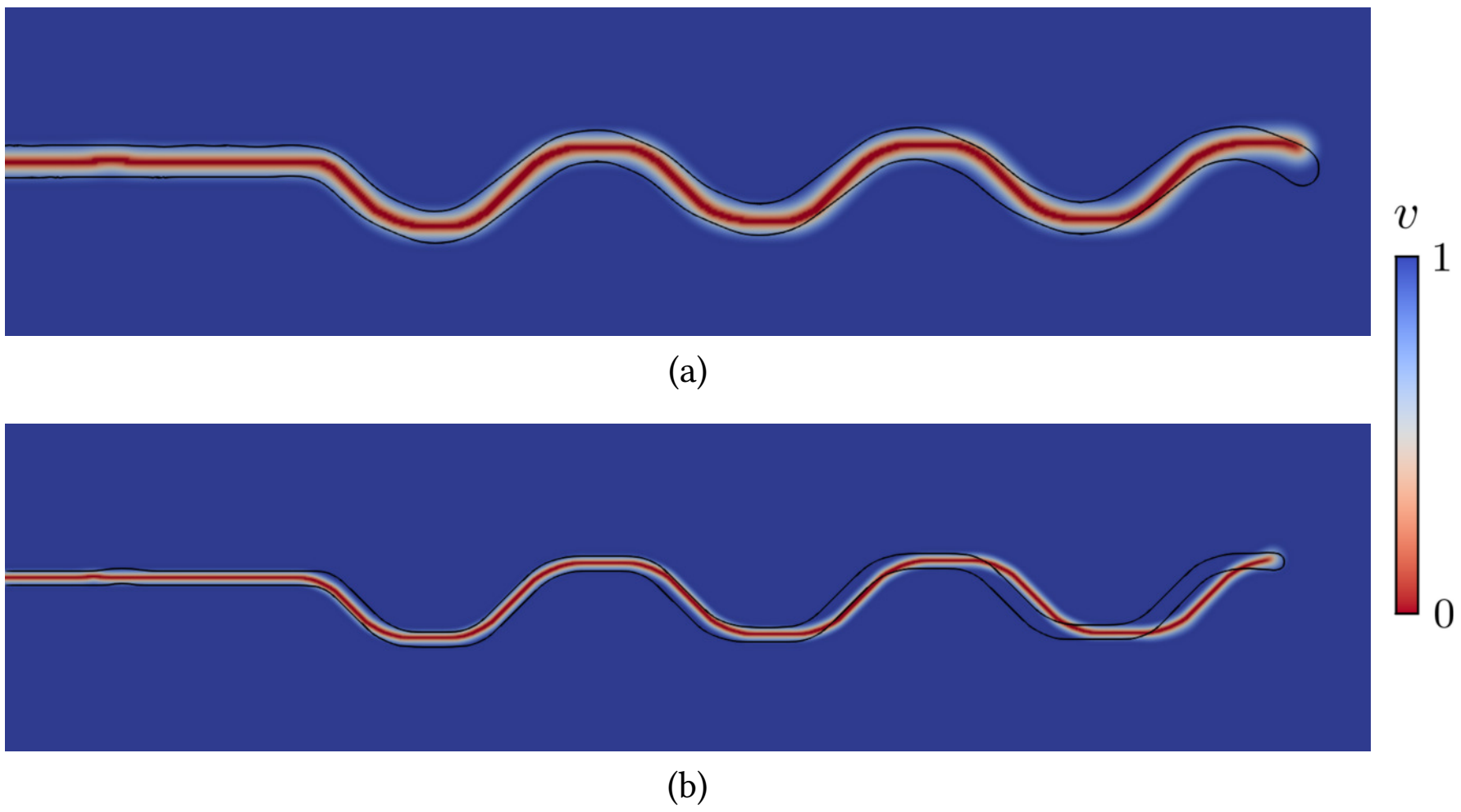

**Figure 10** Simulation results obtained with $\Delta T = 147\,°\mathrm{C}$, $V = 0.4$ mm/s. (a) Results obtained with different element types: The background phase field was obtained from quad4 elements, while the superimposed black contour was extracted from $v = 0.9$ for a comparable mesh that used tri3 elements. (b) Results obtained using quad4 elements, but with different regularization lengths $\ell$: the background phase field was obtained using a regularization length that was 50 % smaller than that indicated by the black contour.

## 3.2 Thin disk fracture under thermal shock

In this section, we examine the capability of the complete model to reproduce and explain the fracture patterns reported for a series of radiative heating experiments on ceramic disks. The experimental setup was originally designed by Awaji et al. (1993); Awaji (1999) and later improved by Honda, Suzuki, et al. (2002); Honda, Nishikawa, et al. (2002); Honda et al. (2009). These experiments were proposed as an alternative to conventional quenching as a thermal shock testing method, as they eliminate the need for a coolant medium. Cracks in the disks nucleate in response to stress states that are spatially fairly uniform or from localized stress concentrations at V-notches.

The series of experiments conducted by Honda, Suzuki, et al. (2002); Honda, Nishikawa, et al. (2002); Honda et al. (2009) involved variations in material type, disk dimensions, and initial temperature. In this work, we focus on the subset of experiments performed on alumina disks with a radius $r_o$ of 15 mm, a thickness of 1 mm, and initially at room temperature. In these experiments, the central surface area (with a radius of 7.5 mm) was coated with graphite to enhance infrared radiation absorption. Two types of samples were tested: intact disks and notched disks with a notch length of 2 mm. A detailed schematic of the sample geometry is provided in Figure 11(a). Each disk was positioned in a custom-designed instrument, shown in Figure 11(b). Upon removal of the shutters that blocked infrared radiation on both sides, the coated region of the disk was rapidly heated. The resulting temperature gradient induced thermo-mechanical stresses that were sufficient to cause fracture. An acoustic emission detector was used to capture the moment the shutters opened (marking the start of heating) as well as the onset of crack initiation. A pair of thermocouples positioned 10 mm from the specimen center recorded the temperature. At the moment of fracture, the temperatures were reported as 36.5 ± 12.3 °C and 91.5 ± 18 °C for the notched and intact specimens, respectively.

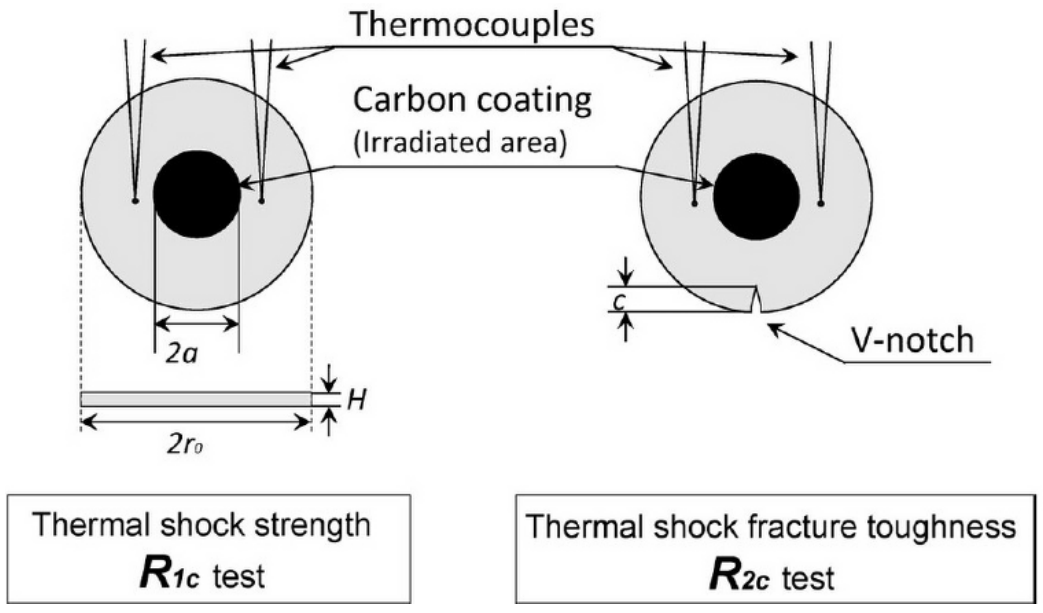

(a)

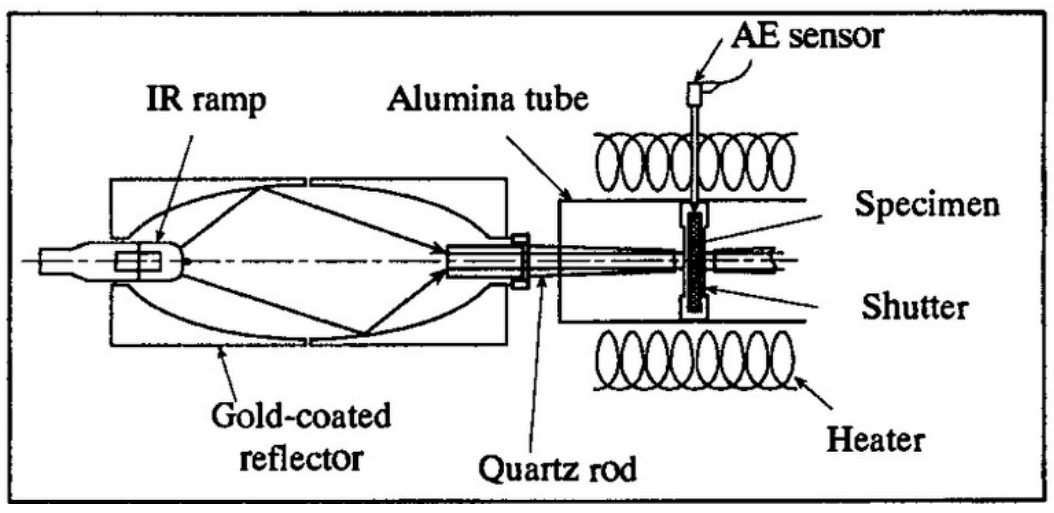

(b)

**Figure 11** Specimen geometries and experimental setup for infrared thermal shock tests on ceramic disks. (a) Schematic diagram of experimental samples used for a fracture "strength test" (left) and a fracture "toughness test" (right) (reproduced from (Honda et al. 2009)). (b) Experimental setup used to induce rapid heating of the central carbon-coated regions, from (Honda, Suzuki, et al. 2002).

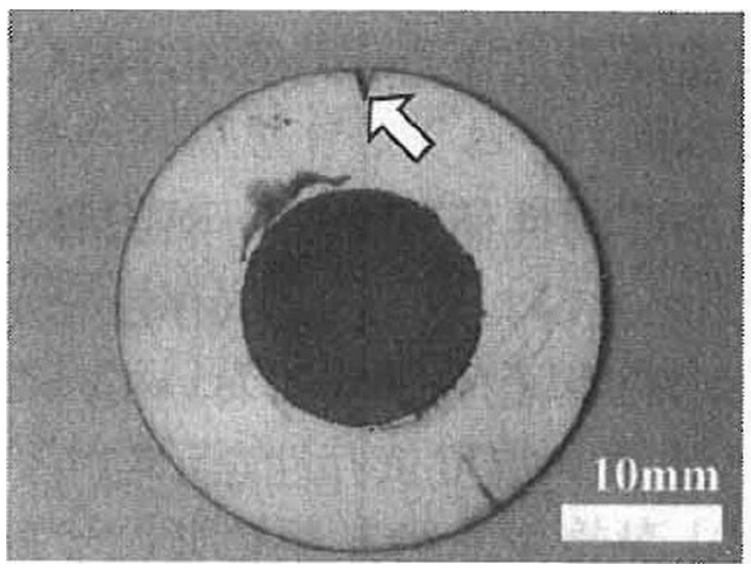

(a)

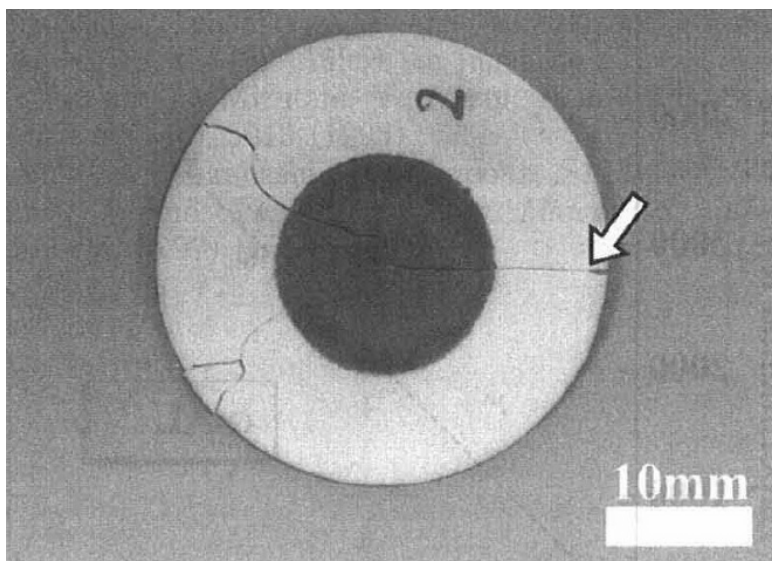

(b)

**Figure 12** Typical crack patterns obtained on the (a) notched disk (Honda, Suzuki, et al. 2002), and (b) on the intact disk (Honda, Nishikawa, et al. 2002). Arrows point to the location of crack initiation in each case.

The two experiments differ in many respects, but perhaps the most interesting concerns the fracture patterns that were observed. For the notched samples, a straight crack propagated from the notch across the entire disk, as shown in Figure 12(a). In contrast, for the intact samples, the crack initiated approximately 2–3 mm from the edge, propagated straight toward the center, and then branched and curved before reaching the opposite edge, as illustrated in Figure 12(b).

Several attempts at reproducing these experiments with phase-field fracture models have been made (Chu et al. 2017; Chen et al. 2024; Li et al. 2023; Mandal et al. 2021; Wang et al. 2018). Notably, in those studies, the notched geometry was often used to simulate both the notched and intact disk crack patterns. In contrast, with the implementation of the complete model, we will demonstrate that the full set of results can be reproduced based on the original experimental configurations.

For the simulation results that follow, several modeling assumptions were invoked. Since the specimens are relatively thin and heated from both sides, the problems are approximated as two-dimensional disks under plane-stress conditions. The material is assumed to be stress-free at room temperature (R.T.) 23 °C. In terms of the thermal loading, heat losses due to the specimen contact with the surrounding air and the fixtures were both neglected. The radiative heating is approximated as a constant, spatially uniform volumetric heat source over the region corresponding to the graphite-coated surface. To reconstruct the unreported thermal loading, thermal simulations were conducted over a range of values for the magnitude $\gamma$ of the volumetric heat source. Although this thermal-load identification was not exhaustive, a value of $\gamma = 0.3\,\mathrm{GW\,m^{-3}}$ was found to give a temperature at the thermocouple location that was within the experimental range observed for the intact disk.

| property/parameter | symbol | unit | R.T. | 200 °C | 400 °C | 600 °C | 800 °C |
|---|---|---|---|---|---|---|---|
| Young's modulus | $E$ | GPa | 380 | | | | |
| Poisson's ratio | $\nu$ | | 0.22 | | | | |
| Density | $\rho$ | $10^3\,\mathrm{kg\,m^{-3}}$ | 3.90 | | | | |
| Fracture toughness | $G_c$ | $\mathrm{N\,m^{-1}}$ | 58 | | | | |
| Thermal conductivity | $\kappa$ | $\mathrm{W\,m^{-1}\,K^{-1}}$ | 27.7 | 21.0 | 14.8 | 10.2 | 9.0 |
| Specific heat | $c_p$ | $\mathrm{J\,kg^{-1}\,K^{-1}}$ | 888 | 962 | 1116 | 1376 | 1256 |
| CTE | $\alpha$ | $10^{-6}\,\mathrm{K^{-1}}$ | 4.5 | 6.6 | 8.8 | 11.0 | 13.3 |
| Mean uniaxial tensile strength | $\bar{\sigma}_{\mathrm{ts}}$ | MPa | 154 | | | | |
| Compressive strength | $\sigma_{cs}$ | | $10 \times \sigma_{\mathrm{ts}}$ | | | | |

**Table 2** Thermal and mechanical properties of alumina from (Honda, Nishikawa, et al. 2002) and fracture toughness from (Krosaki Harima Corporation 2023).

The material properties used in the model are provided in Table 2. Over the range of temperatures considered, the thermal properties of alumina exhibit significant variation. To account for this, the temperature-dependent properties were fitted to the experimentally reported values. Simulations were performed using a regularization length $\ell$ of 0.15 mm and an adaptively refined mesh, with the finest mesh size set to $h = \ell/5$. Relative to the bulk wave speeds, crack growth proceeds very rapidly in these problems, and so inertial effects were included. The initial time step was set to 0.05 s and employed until the moment of fracture, at which point it was reduced to $10^{-7}$ s to resolve the dynamic crack propagation. For the notched disk case only, the symmetry of the problem was exploited so that only one half of the domain was explicitly modeled. The outer boundary of the disk was assumed to be traction free, and zero heat flux conditions were prescribed. No displacement boundary conditions were prescribed at any point.

Almost all of the mechanical properties for alumina used in this study were obtained from (Honda, Nishikawa, et al. 2002). The one exception concerns the fracture toughness, $G_c$. For alumina, a range of estimates for the fracture toughness can be found in the literature. The particular value reported by Honda, Nishikawa, et al. (2002) was $27\,\mathrm{N\,m^{-1}}$, which is considerably lower than the $58\,\mathrm{N\,m^{-1}}$ reported by the original manufacturer (Krosaki Harima Corporation 2023), which is what has been used in this work.

For simplicity, we begin by reporting results prior to the onset of fracture. The simulated temperature fields as a function of time and radial position for the notched and intact disks are provided in Figure 13. For the notched case, the plots correspond to a radial line that passes through the notch. In both cases, the fields indicate a large early temperature rise in the center of the specimen, as anticipated. As time progresses, thermal diffusion results in heating throughout the specimen. The thermal fields give rise to stresses in the domain. Plots of the hoop and radial stress components are provided in Figure 13, once again as a function of radial position. As heat rapidly accumulates in the center, the interior of the specimen initially experiences compressive stresses, while the outer region is dominated by tensile hoop stresses.

For the notched disk, the notch acts to concentrate the stress fields. The hoop stress at the notch tip exhibits a singularity, and its finite element approximation is therefore mesh-dependent. We note that crack nucleation from the tip of the notch does not occur immediately, even though the strength of the material is quickly violated.

In contrast, for the intact disk, the stress distribution is highly symmetric with respect to the disk center. In fact, the results shown for the intact disk, Figures 13(d) to 13(f), are representative of any radial line passing through the center of the domain. Of note, prior to the onset of fracture, the maximum tensile hoop stress occurs at a point that is approximately 3mm away from the edge. This point happens to coincide with the location where crack nucleation was reported for the intact case.

Not surprisingly, the differences in the thermo-mechanical fields between the two cases give rise to very different times for crack nucleation. Table 3 provides a comparison between the experimentally recorded and simulated temperatures at the probe location at the moment of fracture. We recall that the magnitude of the volumetric heat source was calibrated to provide a good match to the experimental temperature for the intact case. Importantly, the simulation also yields a temperature at the thermocouple location that is within the range of the experiment for the notched case. In what follows, we detail the ensuing fracture response for the two cases separately.

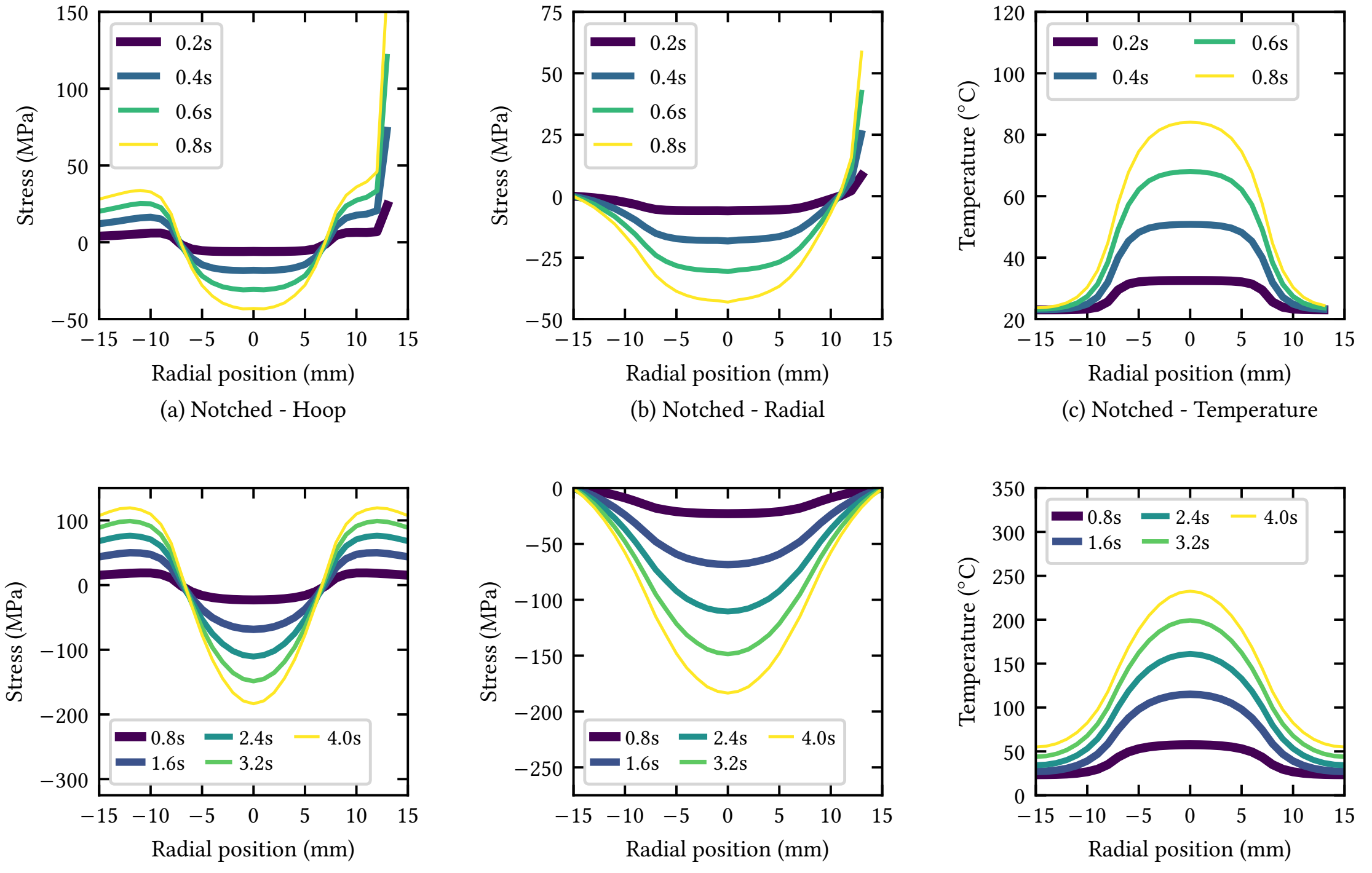

(a) Notched - Hoop (b) Notched - Radial (c) Notched - Temperature

(d) Intact - Hoop (e) Intact - Radial (f) Intact - Temperature

**Figure 13** Simulated stress components (hoop and radial) and temperature profiles along the diameter of the ceramic disks, prior to crack nucleation. The results were obtained by centered heat source with power 0.3 GW m$^{-3}$. The notch tip is located at 13 mm.

**Table 3** Measured temperature at 10 mm from the center, reported in (Honda, Suzuki, et al. 2002), compared with the simulated temperature, at an instant in time immediately prior to fracture.

| Sample type | Notched | Intact |
|---|---|---|
| Experiment | 36.5 ± 12.3 °C | 91.5 ± 18 °C |
| Simulation | 37 °C | 101 °C |

### 3.2.1 Notched disk fracture

In terms of the fracture results, we begin by presenting the simpler case of the notched specimen. Figures 14(a) to 14(e) provide a series of snapshots of the first principal stress. As previously noted, the notch creates a stress singularity and the stress field at the tip of the notch quickly exceeds the strength of the material. We recall that violating the strength surface is a necessary condition for crack nucleation, but it is not sufficient. Although the singular stress fields are mesh dependent, the time to crack nucleation was found to be relatively insensitive to the mesh resolution. In fact, crack nucleation from the notch tip occurred at approximately 1.2 s for all simulations performed with the complete model.

The results with the complete model further indicate that crack propagation from the tip of the notch proceeds immediately afterwards, at $t \approx 1229$ ms. The crack propagates straight through the inner region, even though it is under a state of biaxial compression, and it eventually reaches the opposite edge of the disk, severing it into two pieces. These results are consistent with the experimental observations for the crack path, as shown by Figures 14(f) and 14(g). The simulation predicts a total duration for the notched case, from the moment of initial heating through a fully broken state, of approximately 1229.03 ms. We note that the entire crack propagation takes place rather quickly, over a span of only 20 µs. This is accentuated by the time stamps provided in the snapshots shown in Figures 14(b) to 14(e).

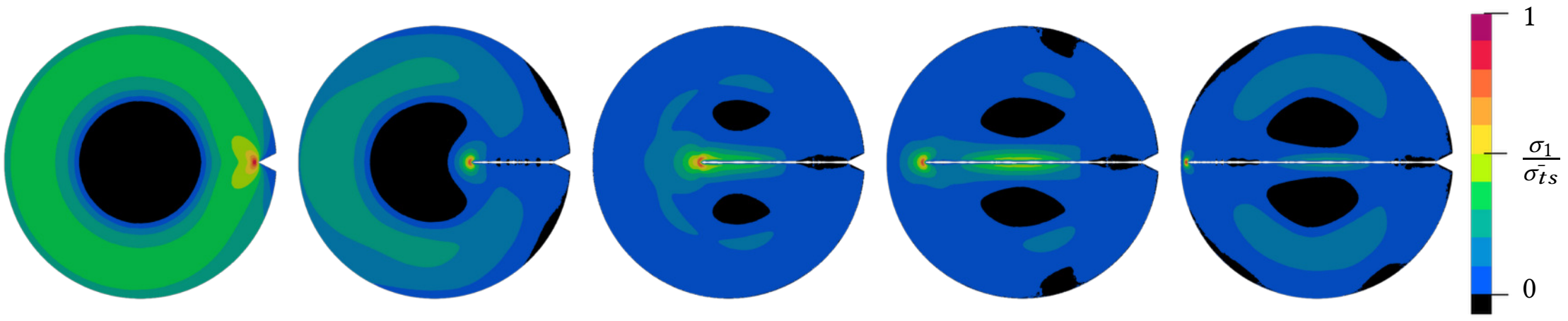

(a) $t_c = 1.2290$s (b) $t_c + 5\mu s$ (c) $t_c + 10\mu s$ (d) $t_c + 15\mu s$ (e) $t_c + 20\mu s$

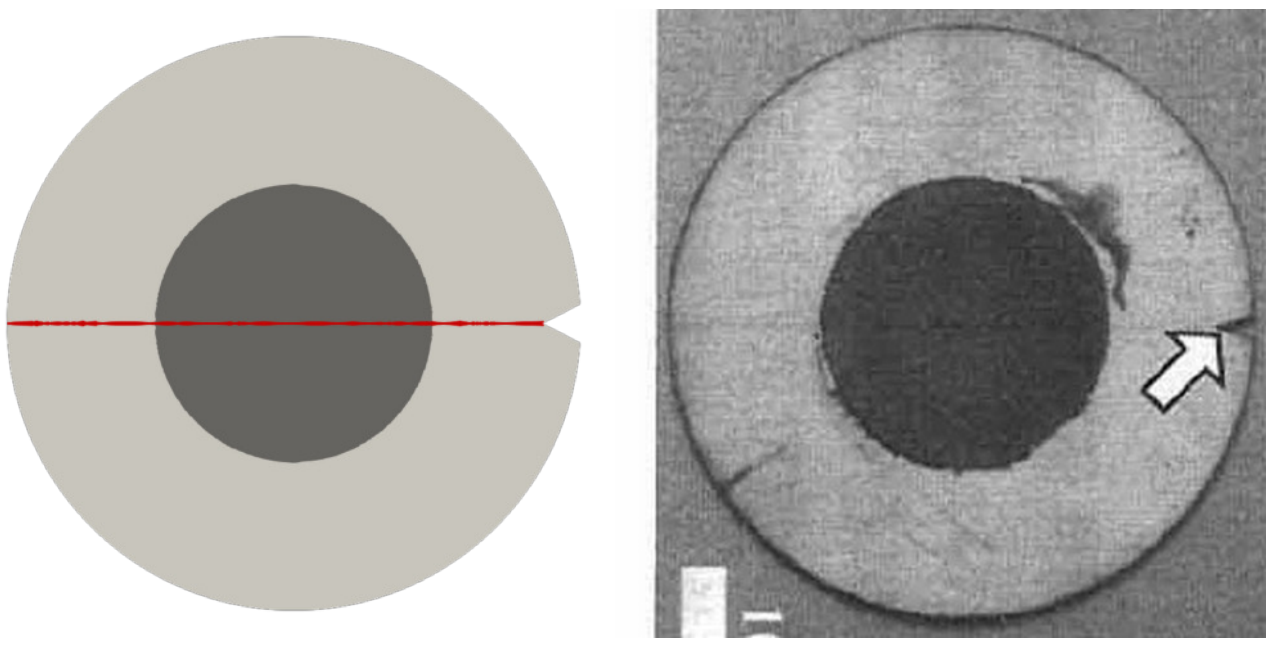

(f) (g)

**Figure 14** Simulation results for the notched disk. (a)-(e) snapshots of the maximum principal stress evolution. (f) the final crack pattern, with the heating area colored in gray and the crack path (filtered by $v < 0.05$) in red. For ease of comparison, the experimental result is shown in (g).

### 3.2.2 Intact disk fracture

We now describe the modeling of the intact disk fracture. As described above, due to the inherent symmetry in this specimen geometry and loading, the state of stress for the intact case only varies in the radial direction. Without any perturbation to the model, the simulation gives a prediction for crack nucleation patterns that are clearly artificial. In what follows, we report results obtained by perturbing the model for the intact case by using a spatially varying strength field[4] and offsetting the center of the heat source by 1 mm. As detailed in Appendix A.1, both perturbations are necessary to yield results that are consistent with the experimental observations for the intact case. Importantly, both of these perturbations can be physically justified.

The crack patterns obtained using the complete phase-field model are presented in Figure 15(f), along with a series of snapshots displaying the maximum principal stress evolution in Figures 15(a) to 15(e). In the intact disk, where no geometric singularity exists to concentrate the stresses, the tensile hoop stress in the exterior gradually increases to exceed the strength surface at $t \approx 3.9$ s, with the area on the right under slightly higher tension. Under the same heating power as the notched case, the longer exposure times lead to a higher level of strain energy that can be dissipated during crack propagation. As the crack enters the compressive region, it branches near the disk center and grows in a winding path towards the opposite side. Consistent with experimental observations, the branch begins within the center region.

Using a different random seed to generate the mosaic perturbation field produced a different final crack pattern, as shown in Figure 15(h), in which one branch underwent secondary branching as it approached the specimen edge.

### 3.2.3 Crack tip velocity profiles

To provide a quantitative view of the crack propagation, the evolution of crack length and crack tip velocity for both disks is shown in Figure 16. The crack tip is tracked using the $v = 0.05$ contour of the phase field. For every three time steps, the average tip velocity is computed over the preceding interval. For the notched case, the crack tip velocity first peaks at ≈ 55 % of the Rayleigh wave speed ($c_R$ = 5750 m/s), then decreases to between $0.2c_R$ and $0.3c_R$ as it propagates through the heated region. Shortly after exiting the heated region, the speed increases to around $0.5c_R$. For the intact disk, by contrast, the tip velocity exhibits a very high peak at the time of nucleation, followed by a rapid decrease shortly after initiation and before entering the heated region. It then quickly arrests inside the compressive zone, before recovering and continuing to propagate at around $0.2c_R$. Notably, crack branching occurs at a tip velocity of $0.2c_R$. This is a much slower speed for dynamic crack branching than is commonly reported in the literature, in which the conventional wisdom is that branching occurs at speeds of approximately $0.6c_R$. We note that in this problem, branching occurs in a region that is under significant compression, in contrast to more commonly studied tensile cases.

[4] In this work, a simple "mosaic" pattern based on the finite element mesh was used to introduce some randomness to the strength field. Alternative constructions that are independent of the mesh can also be used, such as those described in (Zeng et al. 2025).

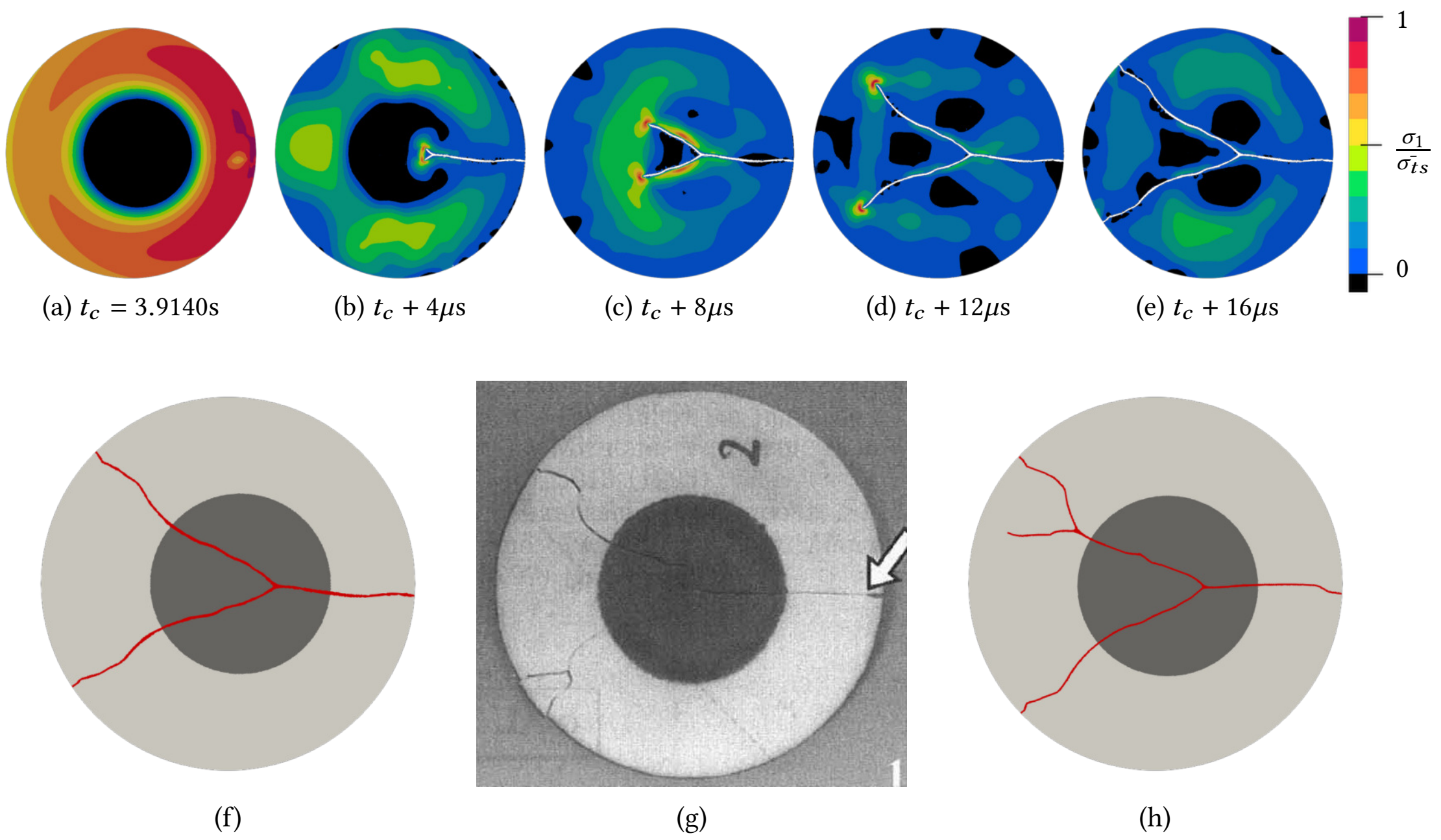

**Figure 15** Simulation results for the intact disk. (a)-(e) snapshots of the maximum principal stress evolution, corresponding to the final crack pattern of (f). (f) and (h) are two final crack patterns, where (h) was obtained with perturbation fields generated using different random seeds compared to (f), with the heating area colored in gray and the crack path (filtered by $v < 0.05$) in red. For ease of comparison, the experimental result is shown in (g).

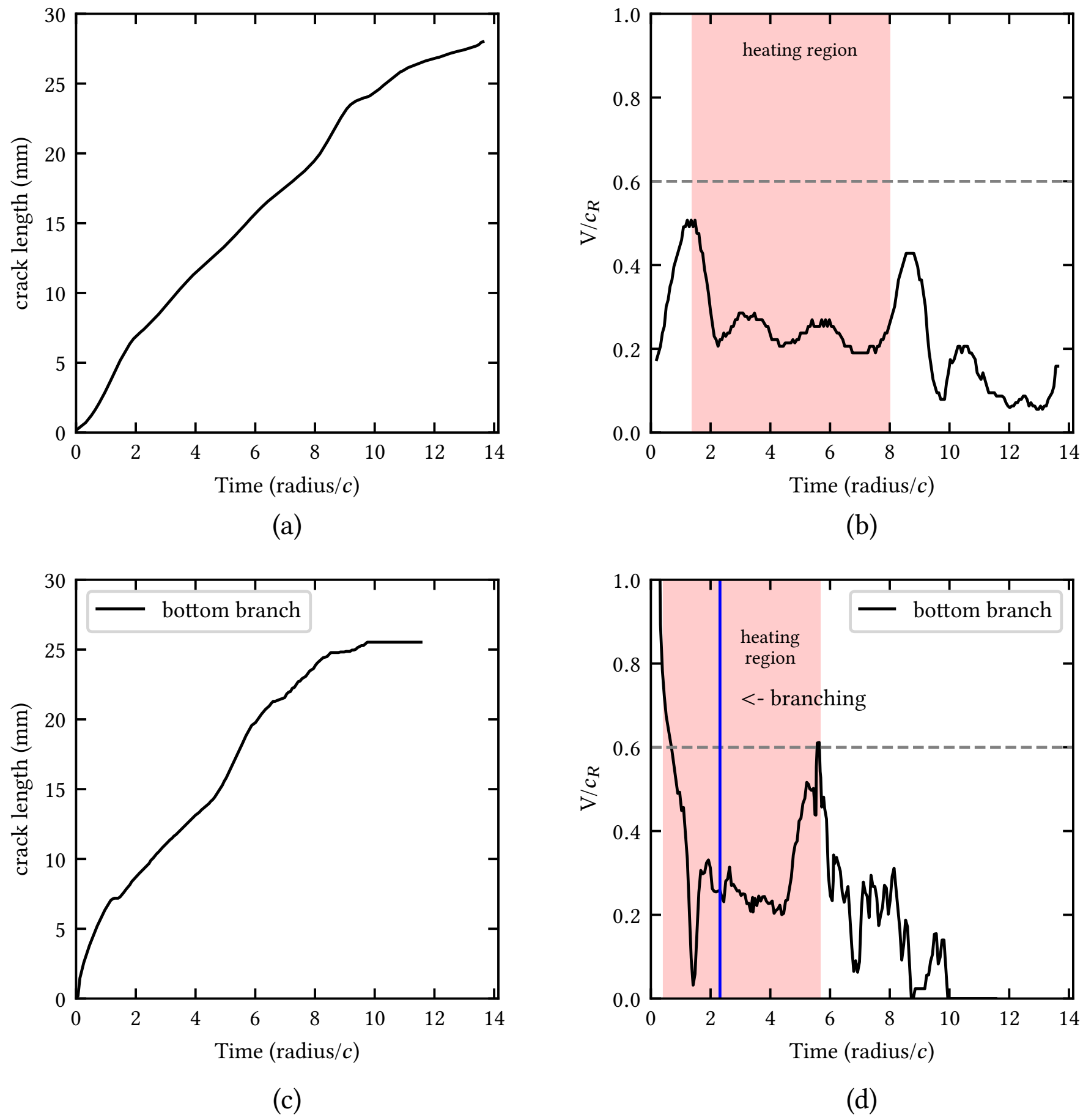

**Figure 16** The evolution of crack length and crack tip velocity, as predicted by the complete model. Simulation results are given in (a)-(b) for the notched disks and (c)-(d) for the intact disks. The time is normalized by the characteristic elastic transit time $R/c$, where $R$ is the disk radius and $c = \sqrt{E/\rho}$ is the characteristic elastic wave speed. The crack tip velocity $V$ is normalized by the Rayleigh wave speed.

We further note that for the notched case, the crack tip velocity was found to be relatively insensitive to the applied heating power. For example, simulations using heating powers of 83 and 50 percent of the estimated experimental power gave rise to nearly identical crack velocity profiles, modulo the different times for nucleation. These findings illustrate the importance of exercising care when calibrating models against experimental observations.

## 3.3 Ceramic fuel fracture under thermo-mechanical loading

In this section, we focus on a series of thermal shock fracture experiments conducted on nuclear fuel pellets, as detailed in a report by Tills (1982). The experiments were carried out at Sandia National Laboratories, as part of an analysis of the Annular Core Research Reactor (ACRR). The ACRR is a pulse reactor that can subject various test objects to a mixed photon and neutron irradiation environment.

During a pulse operation, transient rods are ejected to induce a prompt-supercritical chain reaction of nuclear fission. In this process, a heavy nucleus (U-235) absorbs a neutron and splits, releasing neutrons, gamma rays, and energetic fission fragments. The fast-moving fragments collide with nearby atoms within a few microns of travel and efficiently transfer their kinetic energy into atomic vibrations, thereby heating the fuel.

To test the mechanical integrity of the fuel pellets under extreme pulse operations, a series of in-pile experiments were conducted. Piles of fuel pellets, see Figure 17(a), were shocked under two energy levels of single pulse irradiation and the location of any resulting fractures was recorded.

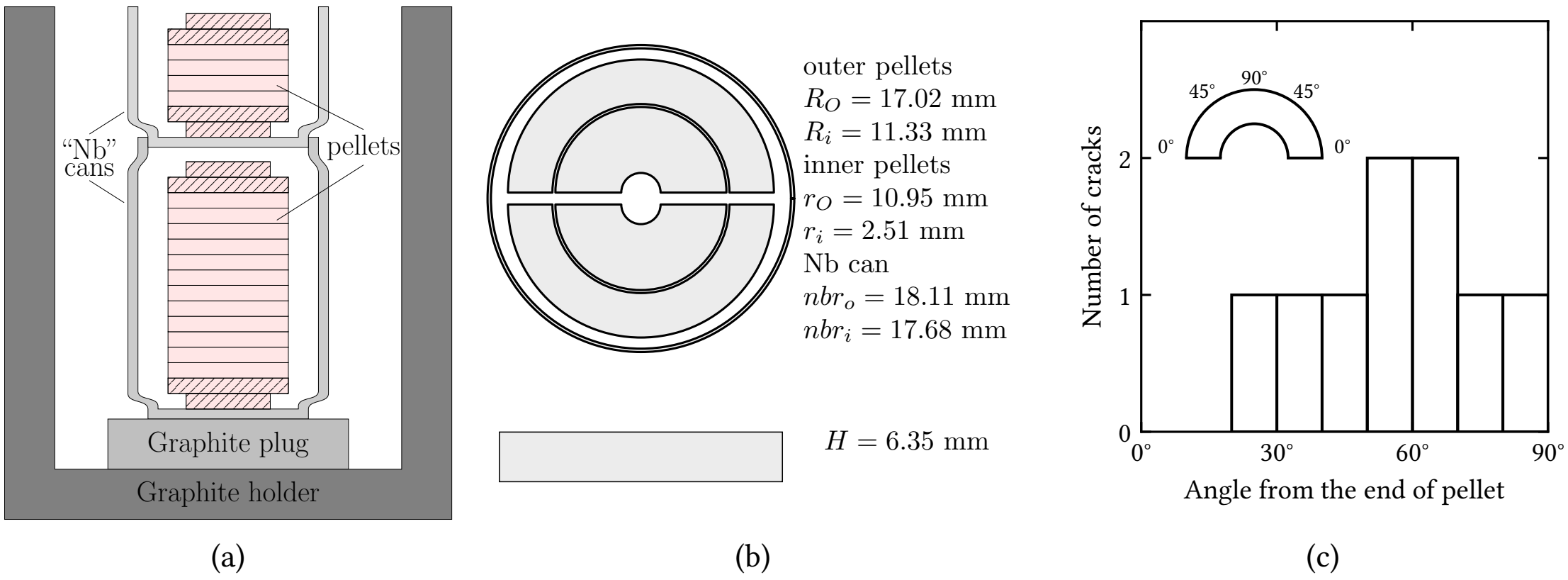

**Figure 17** Single pulse pellet fracture test configuration. (a) Schematic of the containers and specimen stacking. (b) Dimensions of a row of pellets, consisting of a pair of outer pellets and a pair of inner pellets, and the Niobium can. (c) The reported angle distribution of the cracked outer pellets. The shaded pellet layers in (a) were excluded from the statistics due to their unverified fission profile.

The solid fuel tested was BeO–$UO_2$, containing 6.9 % $UO_2$ by volume with 35 % $^{235}$U enrichment. In the experiment, two niobium cans were used to hold 24 total rows of pellets within a graphite holder, as shown in Figure 17(a). Each row consisted of a pair of outer pellets and a pair of inner pellets, as illustrated in Figure 17(b). The entire assembly was positioned at the location of maximum energy deposition in the reactor. Two pulse levels were applied to each set of samples. The lower pulse corresponded to a total reactor energy deposition of 200 MJ, resulting in a theoretical adiabatic outer surface temperature of 1400 °C for the outer pellets. The higher pulse corresponded to 247 MJ of total energy deposition, producing a theoretical 1600 °C outer pellet surface temperature. In what follows, we will refer to each of these cases as the 200 MJ pulse or the 247 MJ pulse, respectively.

As reported in (Tills 1982), all inner pellets remained intact after the pulses. Following the 247 MJ pulse, 25 % of the outer pellets fractured into two pieces. The report also provides the distribution of single-crack positions, expressed as the angle from the edge of the pellet, as shown in Figure 17(c). Due to the symmetry of the pellets, all cracks were reported with angles within the range from 20° to 90°.

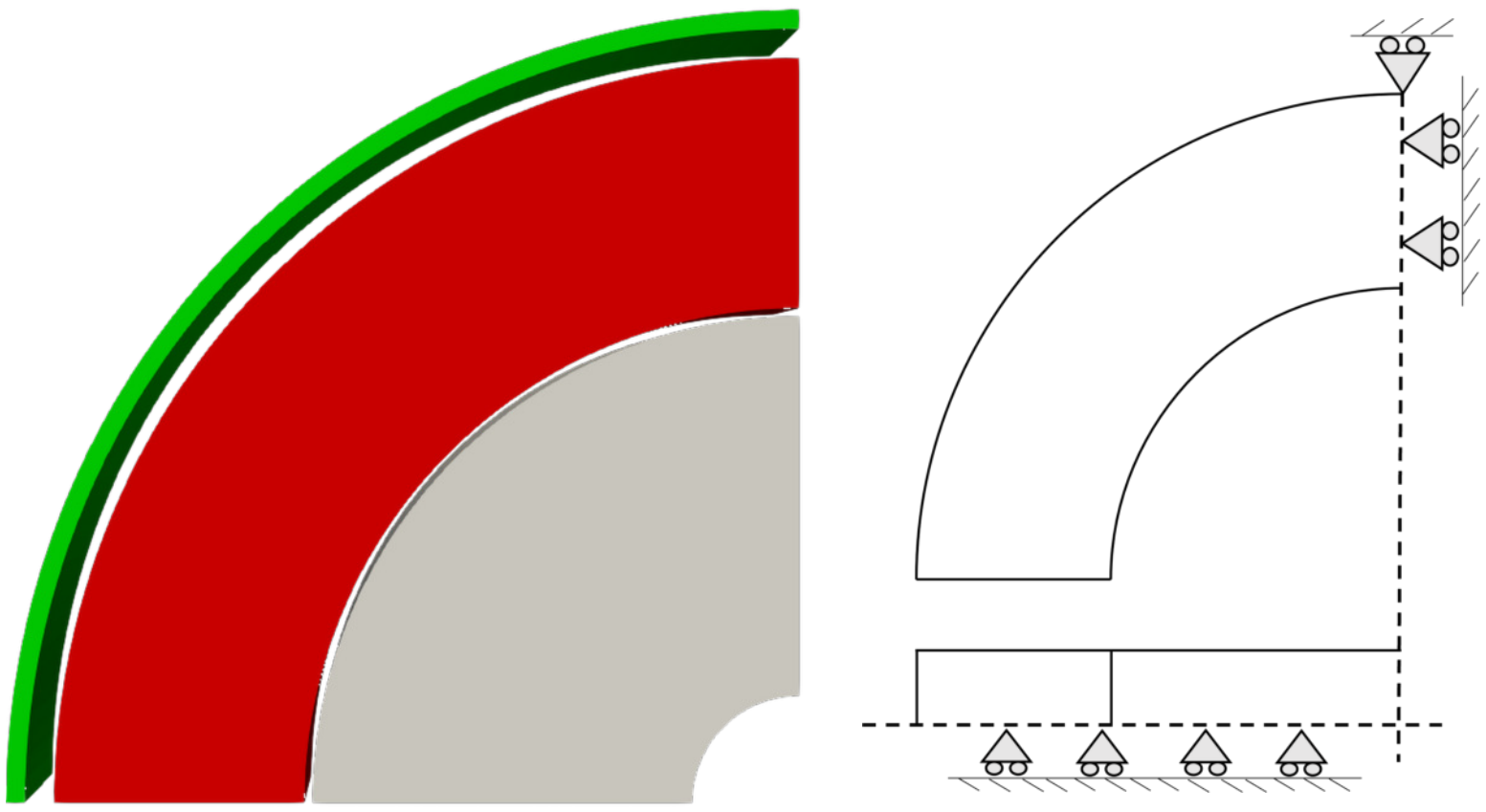

(a) (b)

**Figure 18** Schematic of the modeled geometry. (a) The three components are the Niobium can (green), the outer pellet (red), and the inner pellet (grey). The symmetry boundary condition is illustrated in (b) on the outer pellet for simplicity.

In what follows, we consider a set of inner and outer pellets in the stack, as well as their interaction with the niobium container. The interaction with the pellets above and below is neglected. To reduce computational cost, we exploit geometric symmetry and simulate only a quarter of a pellet layer. Symmetry boundary conditions are imposed on both the displacement and temperature fields along the symmetry planes, allowing only a quarter of the layer (half in height and half in angle) to be modeled explicitly, as illustrated in Figure 18. All components are assumed to be centered in the radial direction, which ensures that no mechanical contact occurs between neighboring parts during the simulation.

The volumetric heat source $Q$ is assumed to be a function of space and time, according to

$$Q = A \times \mathcal{P}(t) \times R(r), \tag{35}$$

where $\mathcal{P}(t)$ is the reactor power trace of a single pulse, and $R(r)$ is the peak/average power deposition along the radius of the pellet. In this expression, $A$ is an amplification factor that is adjusted so that the reactor yields the target total energy deposition during operation. This is taken to be $A = 27\,\mathrm{m}^{-3}$ for the 200 MJ pulse and $A = 26\,\mathrm{m}^{-3}$ for the 247 MJ pulse (Tills 1982).

The radial fission profile provided by Tills (1982) is given by

$$R(r) = 0.79985971 + 0.34048928r - 1.2141774r^2 + 1.939536r^3 - 1.2757009r^4 + 0.32146363r^5 \tag{36}$$

where $R(r)$ is dimensionless and $r$ is the radial coordinate, whose numerical value in centimeters is to be used.

The power trace extracted during the shock (clipped between the initialization and the tail) for the 200 MJ pulse is given by

$$\mathcal{P}(t) = \begin{cases} 10^{35.79} t^{22.59} & 0.05 \leq t \leq 0.075 \\ 10^{-13.1} t^{-20.86} & 0.075 < t \leq 0.1, \end{cases} \tag{37}$$

with the time $t$ measured in seconds. For the 247 MJ pulse, the power trace is given by:

$$\mathcal{P}(t) = \begin{cases} 10^{36.8} t^{23.37} & 0.05 \le t \le 0.075 \\ 10^{-14.31} t^{-22.07} & 0.075 < t \le 0.1. \end{cases} \tag{38}$$

The heat transfer is modeled between each pair of inner and outer surfaces in the radial direction. The heat flux consists of contributions from gap conductance and radiation:

$$q = q_{\text{cond}} + q_{\text{rad}}. \tag{39}$$

For the gap conduction, with $\kappa^{\text{gap}}(T)$ denoting the thermal conductivity of the gas in the gap, $\delta^{\text{gap}}(x)$ the local gap width, and $T_i$ the surface temperatures, the flux is given by

$$q_{\text{cond}} = \frac{\kappa^{\text{gap}}(T)}{\delta^{\text{gap}}(x)}(T_1 - T_2). \tag{40}$$

For the radiative flux, with $f$ denoting the radiation factor and $\sigma = 5.6697 \times 10^{-8}\,\text{W}\,\text{m}^{-2}\,\text{K}^{-4}$ the Stefan–Boltzmann constant, the flux is given by

$$q_{\text{rad}} = f\sigma(T_1^4 - T_2^4). \tag{41}$$

with $\varepsilon_i$ the emissivity coefficient of the surfaces, the radiation factor is

$$f = \left(\frac{1}{\varepsilon_1} + \frac{1}{\varepsilon_2} - 1\right)^{-1}. \tag{42}$$

This radiation model approximates the gap radiation as radiative transfer between two parallel planes, and assumes no energy can escape from the edge of the gap. The emissivity is assumed to be constant, with values for each material listed in Table 4.

| property/parameter | symbol | unit | function ($T$ in °C) |
|---|---|---|---|
| CTE | $\alpha$ | $10^{-6}\,\text{K}^{-1}$ | $-10^{-6}T^2 + 4.288 \times 10^{-3}T + 5.715$ |
| Thermal conductivity | $\kappa$ | $\text{W}\,\text{m}^{-1}\,\text{K}^{-1}$ | $913.9T^{-0.107} - 402.4$ |
| Specific heat | $c_p$ | $\text{J}\,\text{kg}^{-1}\,\text{K}^{-1}$ | $166.86T^{0.2877} + 398.54$ |
| Emissivity | $\varepsilon$ | | 0.37 |
| Density | $\rho$ | $\text{kg}\,\text{m}^{-3}$ | 3573 |
| Poisson ratio | $\nu$ | | 0.2 |
| Young's modulus | $E$ | Pa | $-62.5 \times 10^6 T + 4 \times 10^{11}$ |
| Fracture toughness | $G_c$ | $\text{N}\,\text{m}^{-1}$ | 44[a] |
| Mean uniaxial tensile strength | $\bar{\sigma}_{\text{ts}}$ | MPa | 114[b] |
| Coefficient of variation | | | 0.03 |
| Uniaxial compressive strength | $\sigma_{cs}$ | MPa | $5\sigma_{\text{ts}}$ |
| Helium thermal conductivity | $\kappa^{\text{gap}}$ | $\text{W}\,\text{m}^{-1}\,\text{K}^{-1}$ | $2.5 \times 10^{-4}T + 0.15$ |
| Niobium thermal conductivity | $\kappa^{\text{nb}}$ | $\text{W}\,\text{m}^{-1}\,\text{K}^{-1}$ | $2 \times 10^{-2}T + 50$ |
| Specific heat | $c_p^{\text{nb}}$ | $\text{J}\,\text{kg}^{-1}\,\text{K}^{-1}$ | $5.29 \times 10^{-2}T + 264.61$ |
| Emissivity | $\varepsilon^{\text{nb}}$ | | 0.22 |

[a] We found neither estimate on the tested sample provided in Tills' report nor data on the general BeO-$UO_2$ composite. A reasonable estimate for its range can be based on those of common ceramics (Zirconia, Silicon Nitride, Alumina, Silicon Carbide): $10\,\text{N}\,\text{m}^{-1}$ to $150\,\text{N}\,\text{m}^{-1}$.
[b] The adopted value is explained in the following paragraphs.

**Table 4** Pellet, Niobium and Helium material properties as a function of temperature $T$ (Tills 1982; Petrovic and Haertling 2020; Pelfrey 2019).

In these experiments, the fuel pellets reach very high temperatures. Accordingly, the temperature dependence of various material properties needs to be carefully considered. Expressions for each of the thermo-mechanical material properties and their variation with temperature are provided in Table 4. The CTE, thermal conductivity, specific heat, density, Young's modulus of the pellet and the thermal properties of the Niobium and Helium were obtained from (Tills 1982). The emissivities of the three materials are taken from (Pelfrey 2019). To estimate the tensile strength of the composite pellet material, we begin by noting that the matrix BeO is documented to have $\sigma_{ts} = 124$ MPa (Petrovic and Haertling 2020). We assume that the presence of the $UO_2$ acts to decrease the tensile strength by approximately 8 % to $\sigma_{ts} = 114$ MPa.

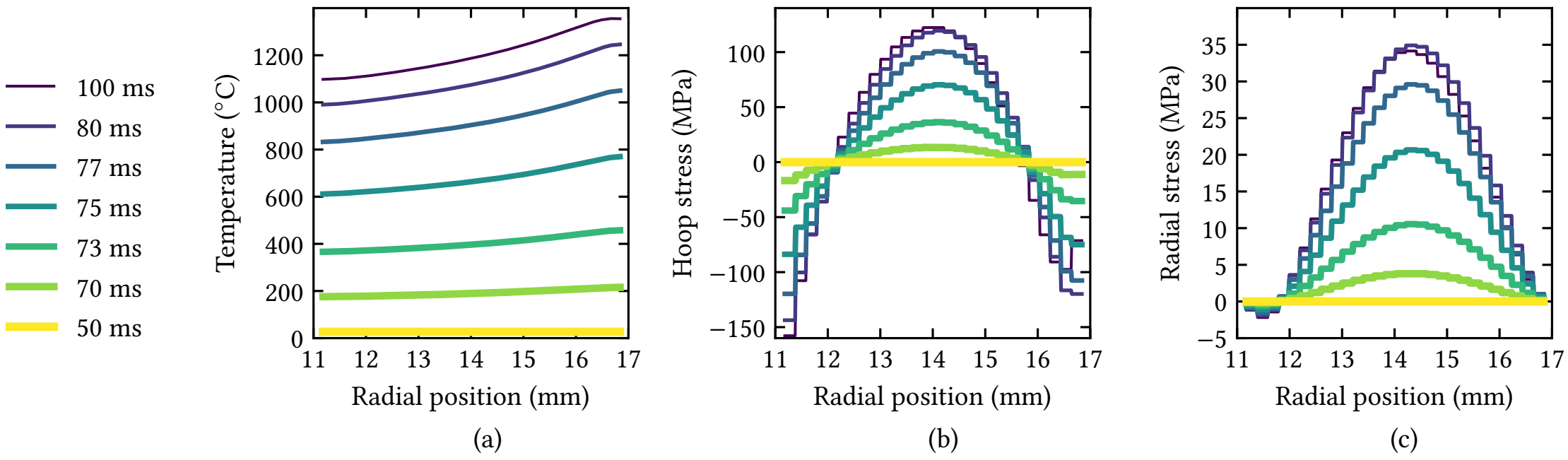

**Figure 19** Thermo-mechanical result of the outer pellet under 247 MJ pulse. (a) Snapshots of temperature distribution along a radial direction on the outer pellet for 247 MJ pulse, (b) the corresponding hoop stresses, and (c) the corresponding radial stresses.

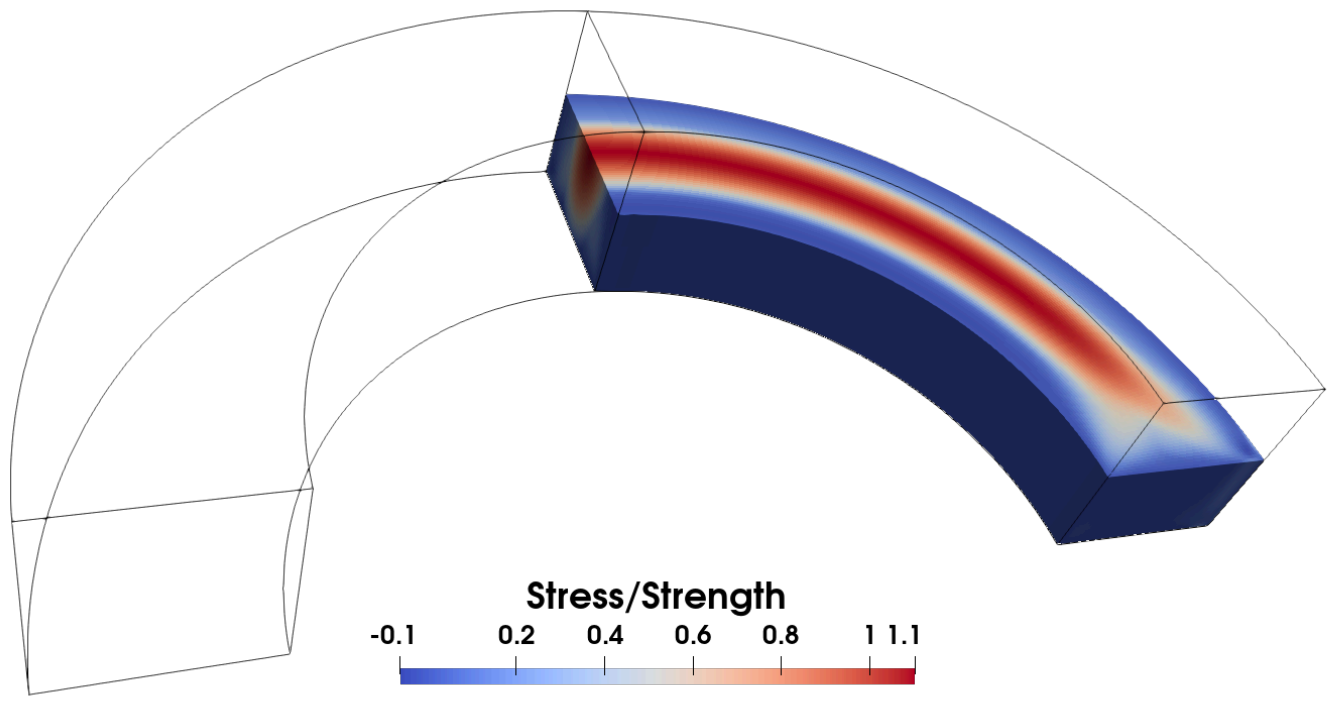

**Figure 20** Maximum principal stress contour of the outer pellet under 247 MJ pulse at 100 ms, normalized by the mean tensile strength.

Our simulations indicate that the outer pellet reaches much higher temperatures than the inner pellet. The larger temperature gradients also give rise to much higher stresses in the outer pellet. For example, Figure 19(a) shows the evolution of the temperature field in the outer pellet for the 247 MJ pulse. The reaction starts at $t = 50$ ms, which gives rise to a rapid increase in the temperature of the outer pellet. In particular, the temperature in the outer pellet increases from room temperature to over 1200 °C by $t = 100$ ms. We note that at each instant in time, the temperature profile monotonically increases from the inner annular surface to the outer surface. A maximum temperature of 1320 °C is obtained on the outer surface of the outer pellet.

The hoop and radial stresses sampled along the 90° centerline of the annulus are shown in Figures 19(b) and 19(c). Along this centerline, the hoop stress is highly tensile in the interior and gradually transitions to compressive near the exterior surface. The radial stress is mostly tensile, peaking in the mid-radius region, but remains generally an order of magnitude smaller than the hoop stress.

We now present the coupled phase-field fracture results. Each numerical sample was assigned a perturbed strength field generated with a different random seed (see Figure 21(a) for one example). A total of 20 samples were simulated at each pulse level. For the 200 MJ pulse, none of the pellets fractured, consistent with the experimental observations[5]. By contrast, 15 outer pellets fractured at the 247 MJ pulse. Among these, 10 samples localized a single crack (see Figure 21(b) for an example), and 5 samples localized two cracks. The cracks were positioned between 40° and 90° from the free end of the pellet. A comparison of the crack angle distributions between the simulations and the experiments is shown in Figure 21(c).

[5] This result is obviously sensitive to our choice of the tensile strength. A much lower value would have given rise to fractures at this smaller pulse level.

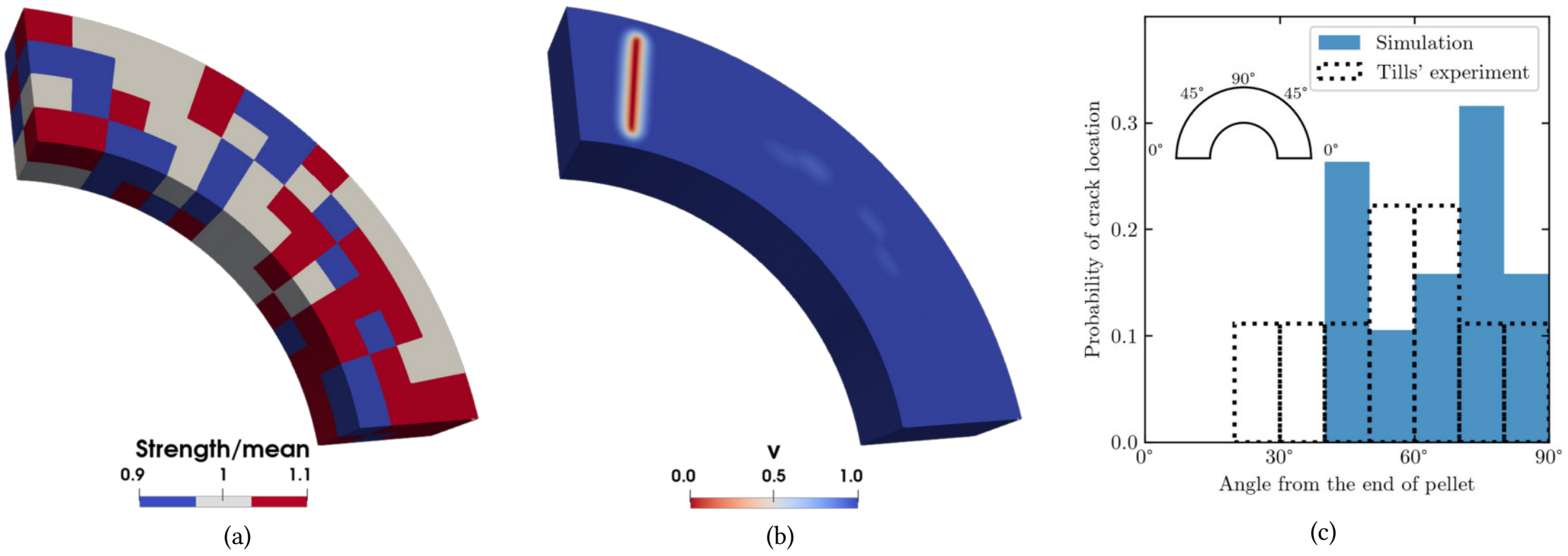

**Figure 21** Strength perturbation, simulated fracture, and crack-location statistics for the outer fuel pellets. (a) An example patch type perturbation field. (b) An example of the fractured pellet, with a through crack penetrating the entire height of the pellet under 247 MJ pulse. (c) Crack distribution comparison between 20 simulated samples with different seeds to generate the random patches and 40 experimental samples.

Although the predicted fractured fraction at 247 MJ is higher than that observed experimentally, the simulations reproduce key qualitative features of the experimental response, including the transition from no fracture at the lower pulse energy to partial fracture at the higher pulse energy. The model also captures variability in fracture outcomes and produces a crack-angle distribution that is broadly consistent with the experimental observations.

Given the uncertainty in the material properties of the fuel pellets, particularly the material strength and fracture toughness, the overall agreement with the experimental trends is encouraging. These results demonstrate the potential of the complete phase-field framework to predict fuel pellet fracture behavior under transient reactor loading, provided that the relevant material properties are characterized more accurately.

## 4 Summary and Concluding Remarks

In this work, we have extended the complete phase-field model of fracture—originally formulated for purely mechanical problems with independent elasticity, fracture toughness, and material strength—to thermo-mechanical loading conditions. The extension retains the key feature of the complete model: a strength surface that is incorporated as an independent constitutive ingredient through an external microforce term in the phase-field evolution equation. Within this framework, thermo-mechanical loading enters only through the elastic strain energy, while the strength surface controls nucleation and the fracture toughness governs propagation. The resulting formulation enables a unified treatment of fracture nucleation and propagation in brittle solids subjected to severe thermal shocks.

The capabilities of the resulting model were examined using three distinct classes of problems that collectively span the canonical fracture scenarios: propagation from a large pre-existing crack, nucleation under nearly uniform stresses, and intermediate states combining localized and distributed stress fields. For the progressive quenching of glass plates, the complete model was found to reproduce the experimentally observed crack patterns and give results that are consistent with a previously established phase diagram. Importantly, parametric studies on the material strength demonstrate that, even for problems traditionally interpreted as governed by Griffith-type crack propagation, the strength surface remains influential.

The infrared-heated alumina disks provide a complementary setting wherein thermo-mechanical loads drive fracture nucleation and branching in the absence of a long pre-existing crack. The employment of the strength surface independently allowed us to first reproduce the intact disk experiment on its original geometry. The competition between strength and energetics that is afforded by the complete model permitted us to explain the two different fracture patterns in this experiment, for the first time.

The fuel-pellet fracture problem illustrated the model's ability to address complex, application-driven thermo-mechanical fracture in nuclear materials. In a statistically perturbed strength field, a moderate pulse (200 MJ) was found to produce no fractures in the outer pellets, whereas a higher pulse (247 MJ) yielded fractured pellets with a distribution of crack angles that was in reasonable agreement with the reported experiments. These results demonstrate that employing perturbed material strength is an effective measure to capture the scatter in failure outcomes at nominally identical loading conditions.

The complete phase-field model offers a framework for predicting brittle fracture in materials subjected to severe thermal shocks. By explicitly separating and independently prescribing the elasticity, fracture toughness, and material strength, the model demonstrated reliable predictions on classical fracture experiment observations and complex nuclear fuel fracture tests.

# A Further discussion on the ceramic disk problem

In this appendix, we present and discuss some supplementary results for the ceramic disk problem.

## A.1 The perturbation of the model

Prior to crack nucleation, a simulation of an intact disk with a perfectly centered heated interior region gives rise to thermo-mechanical fields that are uniform in the circumferential direction. In other words, at any time $t < t_c$, the temperature and stress fields only vary with radial coordinate $r$. The fields along any bisector are indistinguishable from those presented in Figure 13. Without any perturbation to the model, the phase-field patterns that result are those shown in Table A.1(a). In essence, cracks nucleate and localize simultaneously at several locations around the circumference of the disk and briefly propagate before arresting.

Such a pattern is obviously different from what was observed experimentally, wherein crack nucleation occurred at a single location near the disk's perimeter. A reasonable hypothesis is that there is a random aspect to the experimental setup that effectively creates a preferential location for crack nucleation. In what follows, we describe how the introduction of two distinct perturbations is necessary to generate the simulations shown in Figure 15 that provide a strong match to the experimental observations for the intact disk. In particular, we considered: (i) a strength field that varies spatially; and (ii) a slight offset for the center of the heated region.

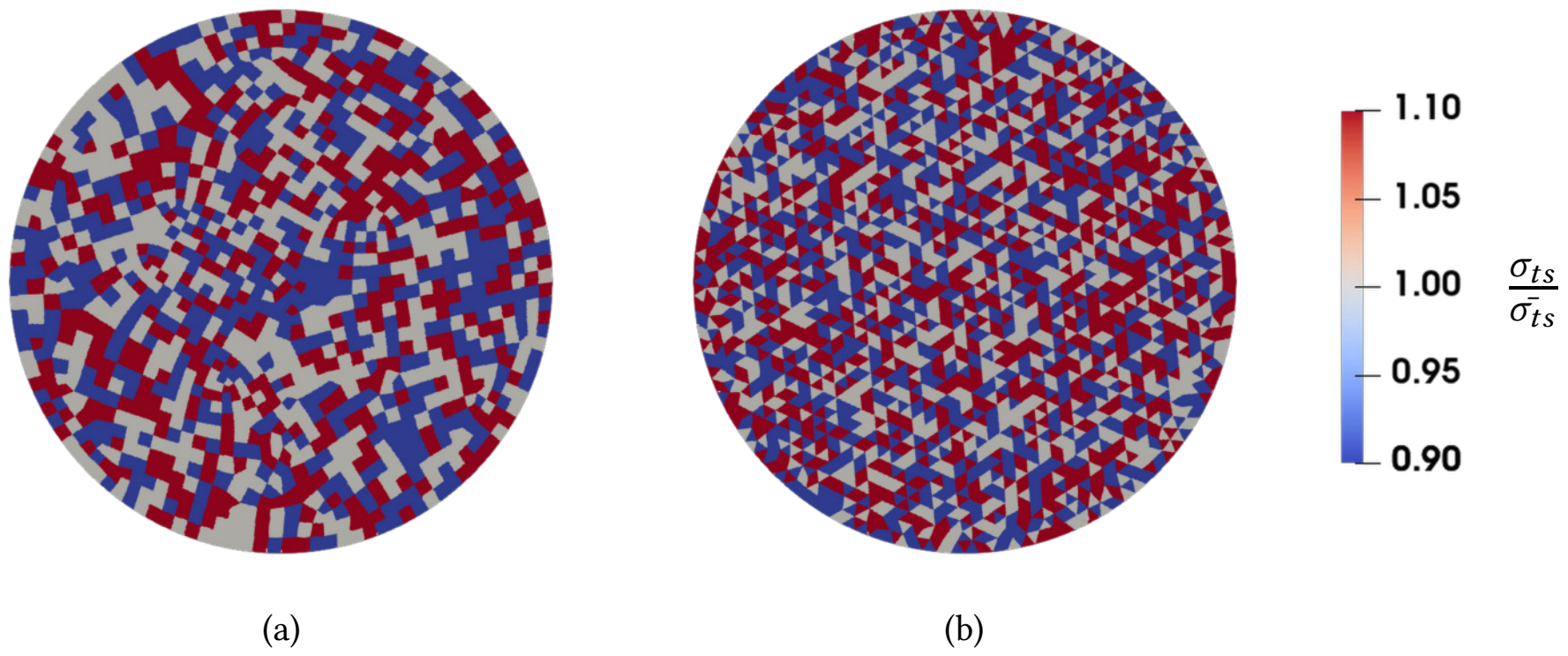

**Figure A.1** Material strength fields with different mosaic types. The field in (a) was sampled on a coarse quadrilateral mesh, and (b) was sampled on a coarse triangular mesh.

We begin by introducing the first perturbation. To perturb the material strength, a mosaic field was introduced as shown in Figures A.1(a) and A.1(b). Patches of elements were randomly assigned material strengths of either 100 %, 110 %, or 90 % of the baseline strength.[6] Each patch spans an area with a characteristic size of roughly $10 \times \ell$. This randomness in material strength promotes asymmetry in crack nucleation around the disk's perimeter. Nevertheless, as shown in Table A.1(c), the results indicate that cracks still nucleate at several locations. By contrast, the experimental observations indicate that fracture typically initiated on one side of the disk, then propagated toward the center and branched toward the opposite edge (see Figure 12). This suggests that some other perturbation likely shifted the system toward a preferred nucleation site.

[6] Adjustments to the baseline strength were effected by modifying both parameters controlling the Drucker-Prager strength surface.

The simplest perturbation to the model that might explain a preference for crack nucleation on one side of the specimen concerns the placement of the heat source. In essence, consider that the coated surface region is not an ideal, centrally located, regular circle. As shown in Figure 11(b), although the exact position of the coated area is difficult to determine, visible irregularities in its shape are evident both within and across samples. This suggests that the level of precision in the coating procedure may have been limited. To incorporate this irregularity into the model, we introduced a slight offset (1 mm) of the heat source center to the right of the domain. Combined with the mosaic perturbation, this adjustment produced a crack pattern closely resembling the experimental observations, as shown in Table A.1(d). For comparison, a simulation performed with only the heat source shift, without material strength perturbation, is provided in Table A.1(b), which resulted in multiple cracks nucleating near the boundary before they coalesced into one that eventually propagated and branched.

| | centered heating | offset heating |
|---|---|---|
| homogeneous strength | 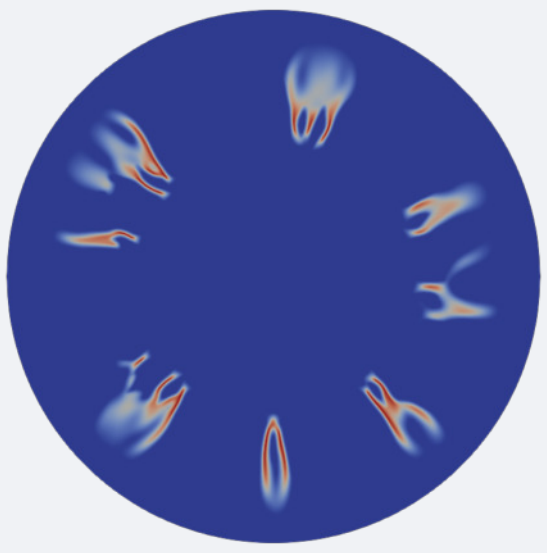 (a) | 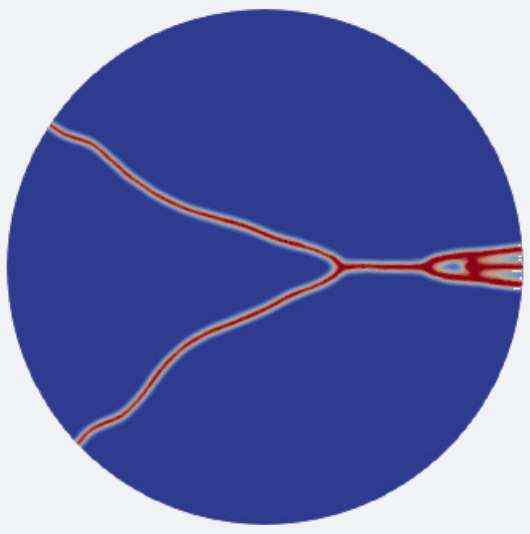 (b) |
| perturbed strength | 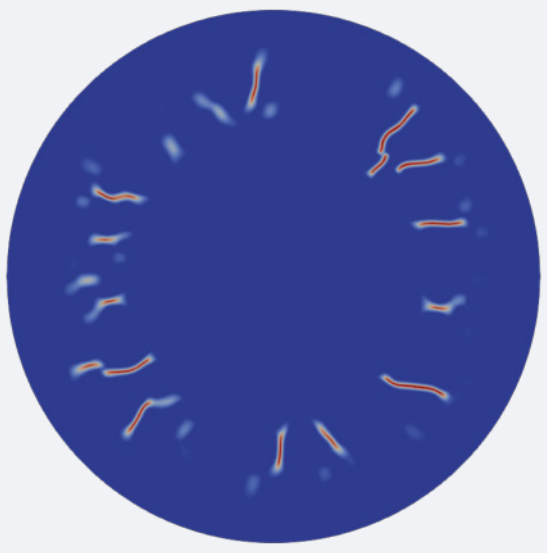 (c) | 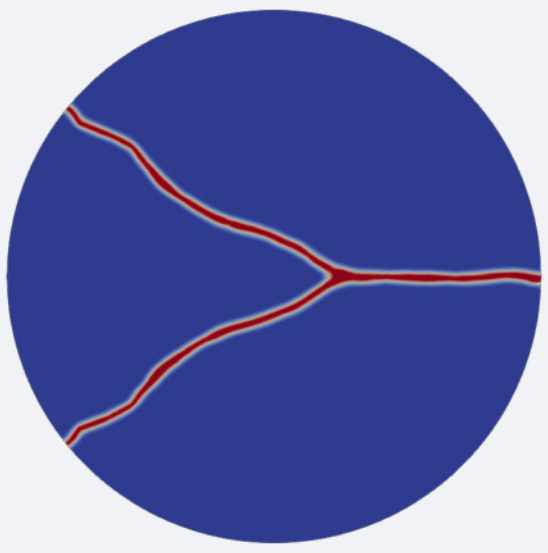 (d) |

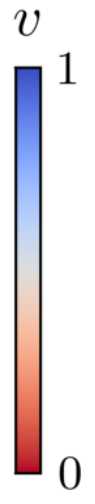

**Table A.1** Fracture patterns obtained by combinations of different heat source locations and material strength fields.

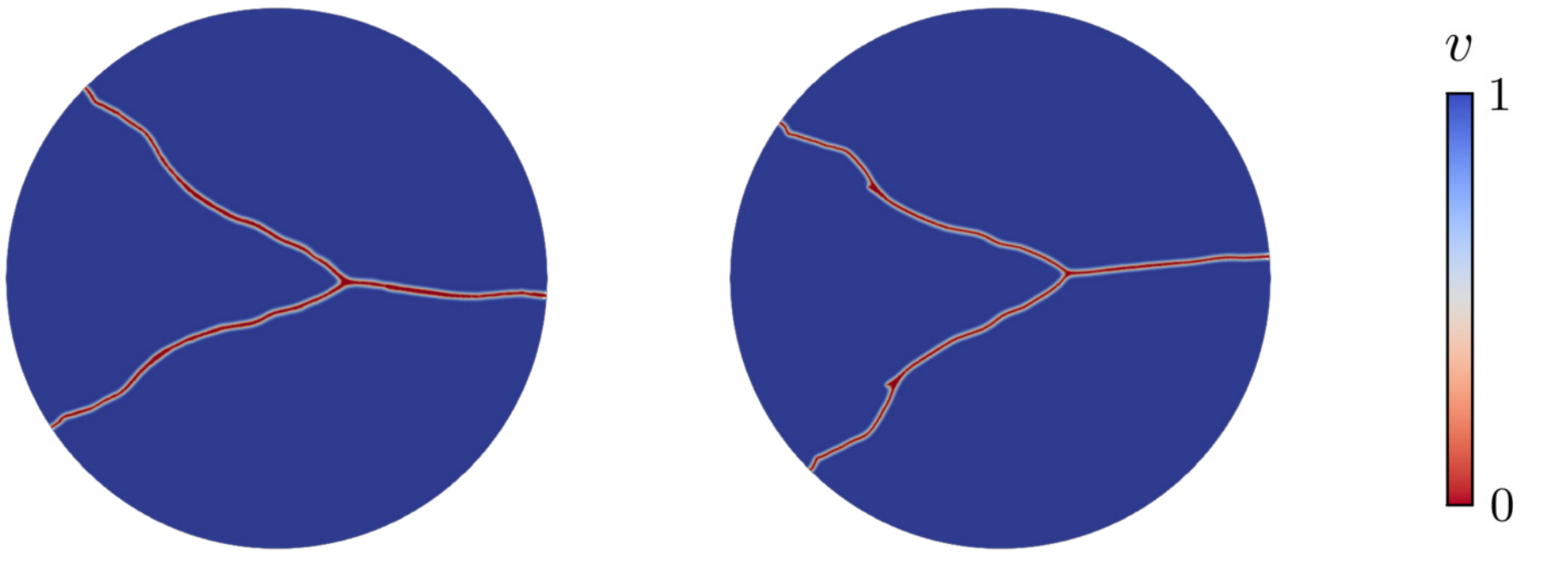

(a) (b)

**Figure A.2** Final crack patterns obtained with strength perturbations from two different mosaic fields: (a) was obtained with a quadrilateral mesh based mosaic field, while (b) was obtained with a triangular mesh based mosaic field. The corresponding perturbation fields are given in Figure A.1.

To conclude this section of the Appendix, we note that one might assume that the construction of the mosaic fields that begin with a coarse mesh might give rise to "mesh-dependent" results. Figures A.2(a) and A.2(b) provide the final crack patterns obtained using the two different mosaic fields from Figures A.1(a) and A.1(b), as well as the shift in the heat source location. The resultant crack patterns present no significant qualitative sensitivity to the distribution of the mosaic random fields. In essence, the primary effect of the mesh mosaic, at least in this case, is to help break the symmetry. Of course, in other situations it is possible that a strong influence of the mesh might be observed. In these cases, the construction of random fields with well-defined correlation lengths, as described in (Zeng et al. 2025), provides an alternative approach that is completely independent of the mesh.

## A.2 Influence from an inappropriate symmetry assumption

The fracture patterns for the intact case, both in this Appendix and in Section 3.2 were obtained using a full disk geometry. In other words, symmetry conditions were not employed to reduce the computational cost by, for example, simulating the response using only the top half of the domain. This is because our simulations indicated that the ensuing crack pattern begins to oscillate slightly about the horizontal axis prior to branching, effectively breaking the symmetry of the problem. By contrast, simulations using only half the domain obviously preclude such an oscillation by construction and were found to compare less favorably to the experiments.

When the whole disk is explicitly modeled, the initial crack propagation on the right side of the disk exhibits a wavy pattern, as shown in Figure A.3(a). Qualitatively, this transition from straight to oscillatory propagation is consistent with the behavior in other problems, such as the progressive quenching of glass plates as discussed in Section 3.1. Importantly, an assumption of symmetry in the fracture response based solely on the symmetric geometry and loading conditions may miss a wavy propagation phase. As shown in Figure A.3(b), when only half of the geometry is modeled with symmetric boundary conditions, a much earlier branching replaces the wavy behavior. One might conclude that this is a problem in which crack path oscillation results rather than early branching because the former is energetically favorable.

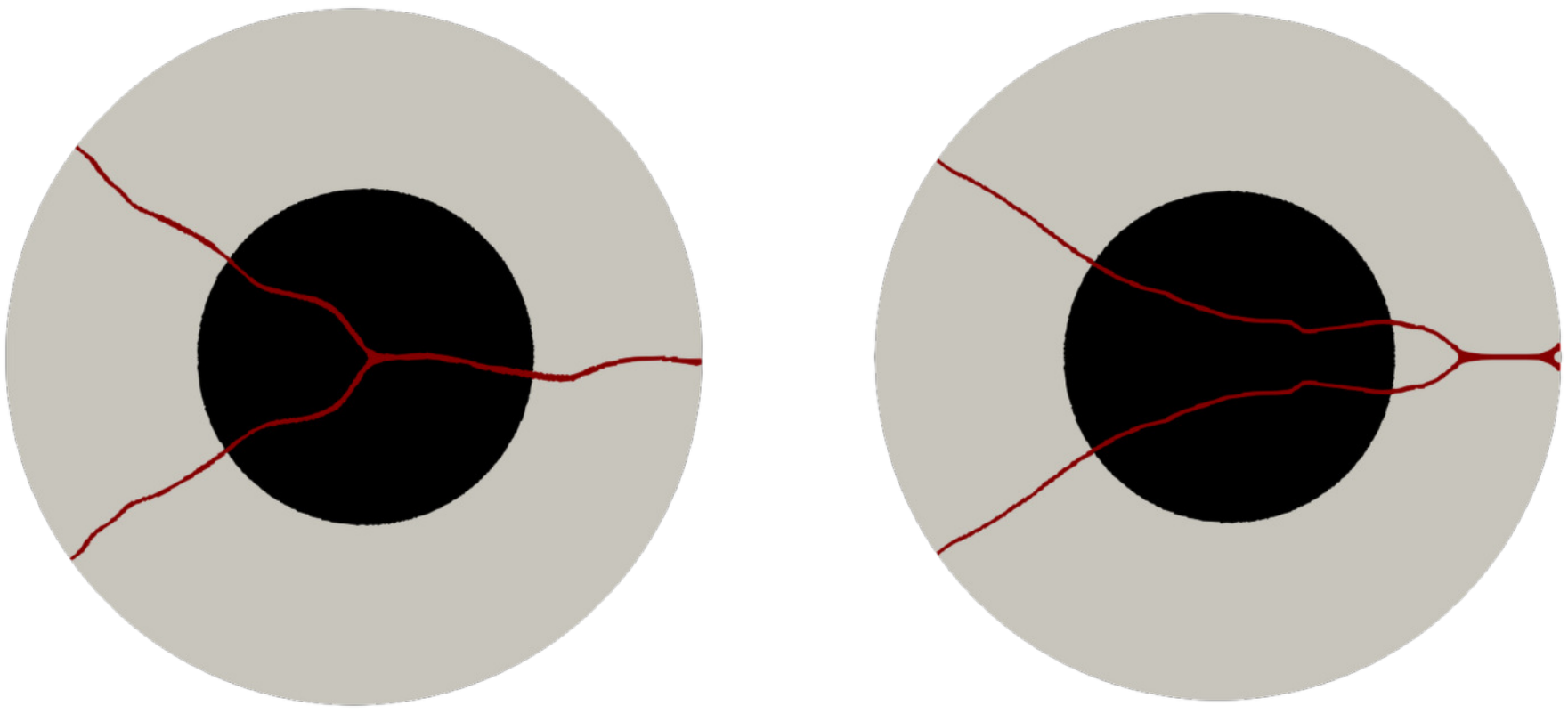

(a) Whole disk simulation (b) Half disk simulation

**Figure A.3** Simulated fracture pattern on the intact disk (a) without and (b) with symmetry conditions imposed on the mid plane.

**Authors' contributions** Bo Zeng: Conceptualization, Methodology, Software, Investigation, Formal analysis, Visualization, Writing – original draft, Writing – review & editing. John E. Dolbow: Conceptualization, Methodology, Supervision, Funding acquisition, Writing – review & editing.

**Supplementary Material** The dataset available at the permalink 10.5281/zenodo.23532161 contains the raw data, simulation outputs, and plotting scripts used to produce the results of Section 3 of the paper. This Open Data approach will allow reproducibility and an easy analysis by another research group.

**Funding** This work was supported by the U.S. National Science Foundation under award CMMI-2132551. The authors are grateful for the support.

**Competing interests** The authors declare that they have no competing interests.

**Journal's Note** JTCAM remains neutral with regard to the content of the publication and institutional affiliations.